\documentclass[twocolumn]{aastex63}

\usepackage{tablefootnote}
\usepackage{amsmath}
\usepackage[dvipsnames]{xcolor}

\usepackage{savesym}
\savesymbol{tablenum}
\usepackage{siunitx}
\restoresymbol{SIX}{tablenum}

\received{Jul 28, 2021}
\revised{Oct 25, 2021}
\accepted{Oct 31, 2021}
\submitjournal{\aj}

\shorttitle{Thermal Phase Variations of XO-3${\rm b}$}

\graphicspath{{./}{figures/}}

\begin{document}

\title{Thermal Phase Curves of XO-3\lowercase{b}:\\ an Eccentric Hot Jupiter at the Deuterium Burning Limit}

\correspondingauthor{Lisa Dang}
\email{lisa.dang@physics.mcgill.ca}

\author[0000-0003-4987-6591]{Lisa Dang}
\affiliation{Department of Physics, McGill University, 3600 University St, Montr\'eal, QC H3A 2T8, Canada}
\affiliation{Universit\'e de Montr\'eal, Institut de Recherche sur les Exoplan\`etes, 1375 Ave.Th\'er\`ese-Lavoie-Roux, Montr\'eal, QC H2V 0B3}

\author[0000-0003-4177-2149]{Taylor J. Bell}
\affiliation{Bay Area Environmental Research Institute, NASA Ames Research Center Moffett Field CA 94035}

\author[0000-0001-6129-5699]{Nicolas B. Cowan}
\affiliation{Department of Physics, McGill University, 3600 University St, Montr\'eal, QC H3A 2T8, Canada}
\affiliation{Universit\'e de Montr\'eal, Institut de Recherche sur les Exoplan\`etes, 1375 Ave.Th\'er\`ese-Lavoie-Roux, Montr\'eal, QC H2V 0B3}
\affiliation{Department of Earth and Planetary Sciences, McGill University, 3450 rue University, Montr\'eal, QC H3A 2A7, Canada}
\affiliation{Space Science Institute, 4765 Walnut St, Suite B, Boulder, CO 80301}

\author[0000-0002-5113-8558]{Daniel Thorngren}
\affiliation{Universit\'e de Montr\'eal, Institut de Recherche sur les Exoplan\`etes, 1375 Ave.Th\'er\`ese-Lavoie-Roux, Montr\'eal, QC H2V 0B3}

\author{Tiffany Kataria}
\affiliation{NASA Jet Propulsion Laboratory, 4800 Oak Grove Drive, Pasadena, CA 91109, USA}

\author[0000-0002-5375-4725]{Heather A. Knutson}
\affiliation{Division of Geological \& Planetary Sciences, California Institute of Technology, 1200 E California Blvd MC 150-21, Pasadena, CA 91125 USA}

\author{Nikole K. Lewis}
\affiliation{Department of Astronomy and Carl Sagan Institute, Cornell University, 122 Sciences Drive, Ithaca, NY 14853, USA}

\author[0000-0002-3481-9052]{Keivan G. Stassun}
\affiliation{Vanderbilt University, Department of Physics \& Astronomy, 6301 Stevenson Center Lane, Nashville, TN, 37235, USA}

\author{Jonathan J. Fortney}
\affiliation{Department of Astronomy and Astrophysics, University of California, Santa Cruz, CA 95064, USA}

\author[0000-0002-0802-9145]{Eric Agol}
\affiliation{Astronomy Department, University of Washington, Box 351580, Seattle, WA 98195 USA}

\author{Gregory P. Laughlin}
\affiliation{Department of Astronomy, Yale University, 52 Hillhouse Avenue, New Haven, CT 06511-8101, USA}

\author[0000-0002-3099-5024]{Adam Burrows}
\affiliation{Department of Astrophysical Sciences, Princeton University Princeton, NJ 08544 USA}

\author{Karen A. Collins}
\affiliation{Vanderbilt University, Department of Physics \& Astronomy, 6301 Stevenson Center Lane, Nashville, TN, 37235, USA}

\author{Drake Deming}
\affiliation{Department of Astronomy, University of Maryland, College Park, MD 20742-2421 USA}

\author{Diana Jovmir}
\affiliation{Department of Physics, McGill University, 3600 University St, Montr\'eal, QC H3A 2T8, Canada}

\author{Jonathan Langton}
\affiliation{Physics Department, Principia College, 1 Maybeck Place, Elsah, Illinois 62028, USA}

\author{Sara Rastegar}
\affiliation{Department of Earth \& Planetary Sciences, Northwestern University, 2145 Sheridan Road, Evanston, IL 60208, USA}

\author{Adam P. Showman}
\affiliation{Lunar and Planetary Lab, University of Arizona Tucson, AZ 85721-0092 USA}



\begin{abstract}

We report \textit{Spitzer} full-orbit phase observations of the eccentric hot Jupiter XO-3b at 3.6 and 4.5 $\mu$m. Our new eclipse depth measurements of $1770 \pm 180$ ppm at 3.6 $\mu$m and $1610 \pm 70$ ppm at 4.5 $\mu$m show no evidence of the previously reported dayside temperature inversion. We also empirically derive the mass and radius of XO-3b and its host star using Gaia DR3's parallax measurement and find a planetary mass $M_p=11.79 \pm 0.98 ~M_{\rm{Jup}}$ and radius $R_p=1.295 \pm 0.066 ~R_{\rm{Jup}}$. We compare our \textit{Spitzer} observations with multiple atmospheric models to constrain the radiative and advective properties of XO-3b. While the decorrelated 4.5 $\mu$m observations are pristine, the 3.6 $\mu$m phase curve remains polluted with detector systematics due to larger amplitude intrapixel sensitivity variations in this channel. We focus our analysis on the more reliable 4.5 $\mu$m phase curve and fit an energy balance model with solid body rotation to estimate the zonal wind speed and the pressure of the bottom of the mixed layer. Our energy balance model fit suggests an eastward equatorial wind speed of $3.13 ^{+0.26} _{-0.83}$  km/s, an atmospheric mixed layer down to $2.40 ^{+0.92} _{-0.16}$ bar, and Bond albedo of $0.106 ^{+0.008} _{-0.106}$. We assume that the wind speed and mixed layer depth are constant throughout the orbit. We compare our observations with a 1D planet-averaged model predictions at apoapse and periapse and 3D general circulation model (GCM) predictions for XO-3b. We also investigate the inflated radius of XO-3b and find that it would require an unusually large amount of internal heating to explain the observed planetary radius.

\end{abstract}




\section{Introduction} \label{sec:intro}

As a transiting planet orbits around its host star, the apparent brightness of the planet varies as seen by a distant observer. Infrared phase variations reveal a planet's response to spatial, diurnal and seasonal forcing. Short-period planets on circular orbits are subject to strong tidal interaction with their host star and hence are expected to have zero obliquity and to be tidally spun down into synchronous rotation. This means that they don't experience seasons or diurnal forcing, so their atmospheric circulation is driven by the fixed spatial pattern of stellar irradiation and Coriolis forces, resulting in steady-state circulation patterns \citep[e.g.][]{2002A&A...385..166S}. Their thermal phase curves can therefore be translated into a longitudinal thermal map of the planet.

While most hot Jupiters have circular orbits, a few have been found on eccentric orbits. These gas giants are expected to form either in-situ \citep{2000Icar..143....2B} or beyond the snow line in their protoplanetary disk far from their stellar hosts and later migrate inwards via gas disk \citep{1996Natur.380..606L}, planet-planet scattering \citep{1996Sci...274..954R}, secular interaction \citep{2011ApJ...735..109W} or Kozai-Lidov migration \citep{1996Natur.384..619W, 2013MNRAS.431.2155N}. In addition to providing insights into gas giants migration mechanisms, hot Jupiters on eccentric orbits are of particular interest for atmospheric studies. 

Unlike their circular counterparts, eccentric hot Jupiters experience time-variable heating such that their phase curve reflects a balance between incoming flux, heat transport efficiency (a combination of rotation and winds), and time required to radiate away energy. Therefore, the variable stellar irradiation experienced by eccentric hot Jupiters allows us to break the degeneracy between the heat transport and radiative timescales that limits our studies of typical hot Jupiters on circular orbit\citep{2008ApJ...674.1106L, 2010ApJ...712..218I, 2011ApJ...726...82C, 2013ApJ...767...76K}.
In contrast with short-period planets on circular orbits, exoplanets on eccentric orbits present additional challenges when one attempts to retrieve information about their atmosphere from thermal phase observations. In particular, it is difficult to disentangle the flux variation due to the planet's rotation and the change in stellar irradiation. While the dayside of the planet should experience spatial and temporal variability over the course of an orbit, the phase curve should be relatively stable from one orbit to the next \citep{2009ApJ...699..564S, 2010ApJ...720..344L, 2013ApJ...767...76K}.

Thermal phase variations have now been published for more than a dozen gas giants on circular orbits with the \textit{Spitzer} Space Telescope: 
\mbox{CoRoT-2b} \citep[][PID 11073]{2018NatAs...2..220D};
\mbox{HAT-P-7b} \citep[][PID 60021]{2016ApJ...823..122W};
\mbox{HD 149026b} \citep[][PID 60021]{2018AJ....155...83Z};
\mbox{HD 189733b} \citep[][PID 60021]{2012ApJ...754...22K};
\mbox{HD 209458b} \citep[][PID 60021]{2014ApJ...790...53Z};
\mbox{KELT-1b} \citep[][PID 11095]{2019AJ....158..166B};
\mbox{KELT-9b} \citep[][PID 14059]{2020ApJ...888L..15M};
\mbox{KELT-1b} \citep[][PID 11095]{2019AJ....158..166B};
\mbox{KELT-16b} \citep[][PID 14059]{2021MNRAS.504.3316B};
\mbox{MASCARA-1b} \citep[][PID 14059]{2021MNRAS.504.3316B};
\mbox{Qatar-1b} \citep[][PID 13038]{2020AJ....159..225K};
\mbox{WASP-12b} (\citealt{2012ApJ...747...82C}, PID 70060; \citealt{2019MNRAS.489.1995B}, PID 90186);
\mbox{WASP-14b} \citep[][PID 80073]{2015ApJ...811..122W};
\mbox{WASP-18b} \citep[][PID 60185]{2013MNRAS.428.2645M};
\mbox{WASP-19b} \citep[][PID 80073]{2016ApJ...823..122W};
\mbox{WASP-33b} \citep[][PID 80073]{2018AJ....155...83Z};
\mbox{WASP-43b} \citep[][PID 11001]{2017AJ....153...68S}; 
\mbox{WASP-76b} \citep[][PID 13038]{2021arXiv210703349M}; and
\mbox{WASP-103b} \citep[][PID 11099]{2018AJ....156...17K}. The many published \textit{Spitzer} thermal phase curves have also enabled various comparative studies \citep{2018AJ....156...28A, 2019NatAs...3.1092K, 2021MNRAS.504.3316B}. Thermal phase curves have also been observed with the \textit{Hubble} Space Telescope \citep[e.g.][]{2014Sci...346..838S, 2018AJ....156...17K, 2019A&A...625A.136A, 2021A&A...646A..94A} and more recently with the Transiting Exoplanet Survey Satellite \citep[e.g.][]{2020AJ....159..104W, 2021AJ....161..131D, 2021A&A...648A..71V}. In contrast, only 3 exoplanets with an eccentricity greater than 0.15 have published phase curves: HAT-P-2b \citep[][PID 90032]{2013ApJ...766...95L}, GJ 436b \citep[][PID 30129]{2014A&A...572A..73L} and HD 80606b \citep[][PID 60102]{2016ApJ...820L..33D}.

\begin{figure}[!htbp]
\includegraphics[width=0.95\linewidth]{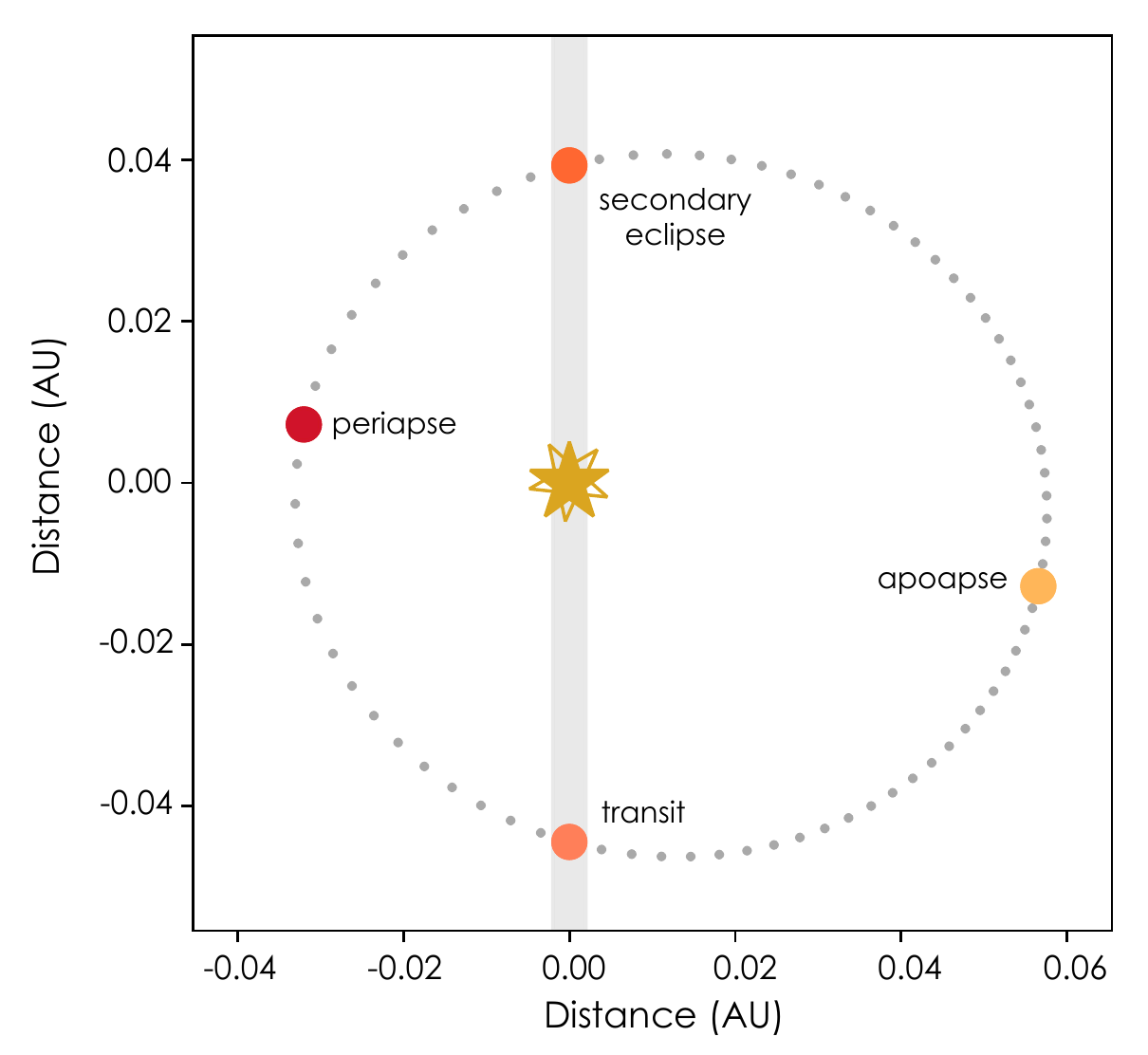}
\caption{Top-view of XO-3b's orbit with parameters from \cite{2014ApJ...794..134W}. The gray dots represent the planet's position at 1 hour intervals and the planet revolves counter-clockwise. The shaded area represents our line of sight (the Earth is off the bottom of the plot). \label{fig:orbit}}
\end{figure}

XO-3b \citep{2008ApJ...677..657J} is an eccentric hot Jupiter ($e=0.2769$) orbiting a F5V star (Figure \ref{fig:orbit}) and is a tantalizing target for follow-up observations. With a mass of $M_p=11.79 \pm 0.98 M_{\rm{Jup}}$, XO-3b provides an important link between giant exoplanets and low-mass brown dwarfs. In addition, its unusually large radius measurements of $R_p = 1.295 \pm 0.066 R_{\rm{jup}}$ is difficult to explain with traditional evolution models for hot Jupiters with an age of $2.82 ^{+0.58} _{-0.82}$ Gyr \citep[e.g.][]{2008ApJ...687.1191L, 2008ApJ...683.1076W}. Finally, the incident stellar flux on XO-3b at periapse is 3.3 times that at apoapse, which could cause large variations in atmospheric temperature, wind speed, chemistry, and clouds.

Naturally, the puzzling system of XO-3 has been the subject of various observational studies. Planets on eccentric orbits are often misaligned with their stellar spin. \cite{2008A&A...488..763H} measured a sky-projected spin-orbit misalignment for the star XO-3 of $\lambda = 70 \pm 15 ^{\circ}$ using SOPHIE observations. This quantity has been revised by \cite{2009ApJ...700..302W} to $\lambda = 37.3 \pm 3.7 ^{\circ}$. \cite{2017MNRAS.472.3871T} obtained ground-based transit observations and suggest that the anomalously large transit depth they measure in \textit{B}-band could be indicative of scattering in its atmosphere. \cite{2010ApJ...711..111M} measured eclipses of XO-3b in the four \textit{Spitzer}/IRAC wavebands to infer the planet’s vertical temperature profile. They found that the dayside of the planet exhibited a temperature inversion (temperature increasing with height over a limited pressure range). \cite{2014ApJ...794..134W} and \cite{2016AJ....152...44I} analyzed 12 secondary eclipses of XO-3b at 4.5 $\mu$m. These measurements favor a greater eclipse depth than reported by \cite{2010ApJ...711..111M}, and hence strengthen the claimed temperature inversion. With the baseline eclipse observations of XO-3b at $4.5~\mu$m extended to 3 years, \cite{2014ApJ...794..134W} place an upper limit on the periastron precession rate of $2.9 \times 10^{-3}$ deg/day. 

In this paper, we present and analyze the full-orbit 3.6 and 4.5 $\mu$m phase curves of XO-3b  obtained with the Spitzer Space Telescope. In addition, we analyze an unpublished 3.6 $\mu$m \textit{Spitzer} secondary eclipse observation. We use these observations to constrain the radiative and dynamical properties of the atmosphere of XO-3b. The observation and data reduction are presented in Section \ref{sec:obs} and the models and methods are described in Sections \ref{sec:model} and \ref{sec:Param}. Our results are presented in Section \ref{sec:results}. Section \ref{sec:conclu} summarizes the main conclusions from our analysis and presents ideas for future work.



\section{Observation and Reduction} \label{sec:obs}
We observed two continuous full-orbit phase curves of XO-3b (PI H.A. Knutson, PID: 90032): one in each of the 3.6 $\mu$m (channel 1) and 4.5 $\mu$m (channel 2) bands of the Infrared Array Camera \citep[IRAC;][]{2004ApJS..154...10F} on board the Spitzer Space Telescope \citep{2004ApJS..154....1W}. The observations at 3.6 $\mu$m and 4.5 $\mu$m were acquired on UT 2013 April 12-16 and UT 2013 May 5-8, respectively. Both observations were scheduled to start approximately 5 hours before the start of a secondary eclipse and end approximately 2 hours after the subsequent secondary eclipse. Due to long-term drift of the spacecraft pointing, the telescope was repositioned approximately
every 12 hr in order to re-center the target. Hence the observations at each waveband were separated into 9 Astronomical Observation Requests (AORs). We used the sub-array mode with a 2 second frame time (effective exposure time of 1.92 s) for 3.56 days in each waveband which yielded a total of 2400 datacubes in each channel. Every datacube consists of 64 frames of 32$\times$32 pixels (39''$\times$39'') resulting in a total of 153,600 images in each waveband. 

As explained later, the 3.6 $\mu$m phase observations exhibit strong detector systematics during one of the secondary eclipse that biases the eclipse depth measurements. To better constrain the 3.6 $\mu$m eclipse depth, we also analyze the eclipse portion of XO-3b's 3.6 $\mu$m partial phase curve (PI: P. Machalek, PID 60058). The 3.6 $\mu$m phase observations were acquired on UT 2010 March 21-23. Again, the sub-array mode with a 2 second frame time (effective exposure time of 1.92 s) was used and a total 419 datacubes were included in our analysis.

\begin{figure}[!htbp]
\includegraphics[width=\linewidth]{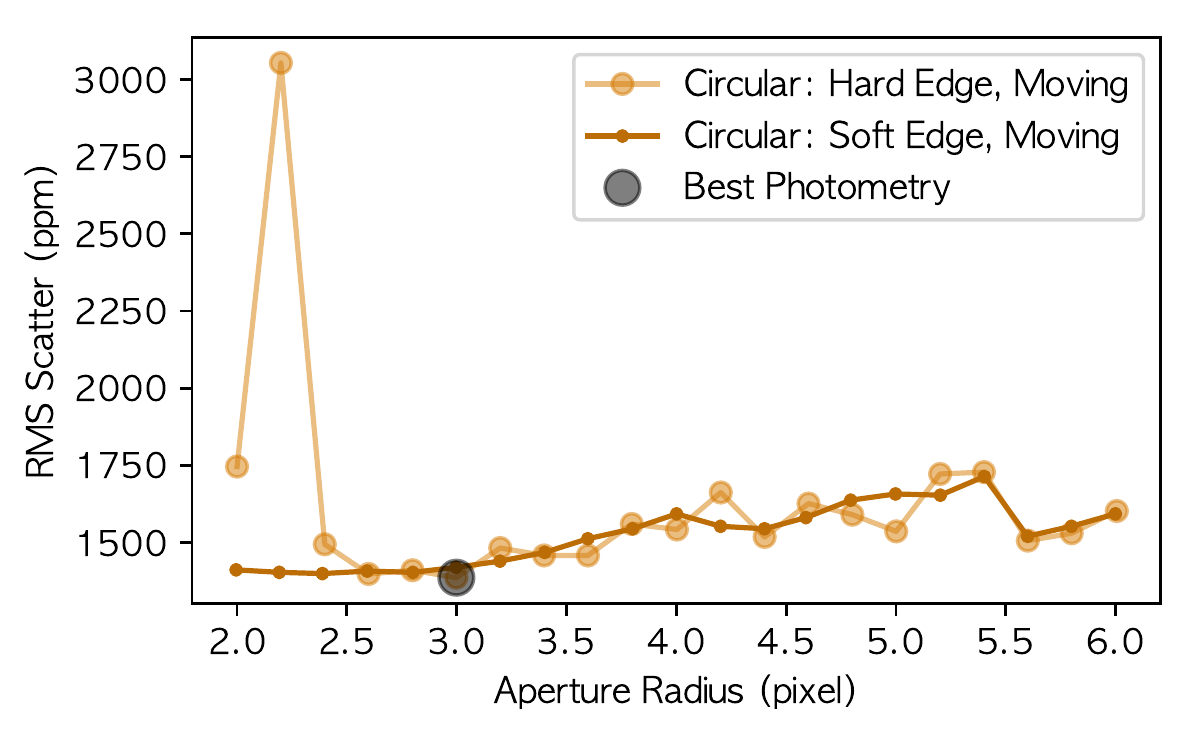}\\
\includegraphics[width=\linewidth]{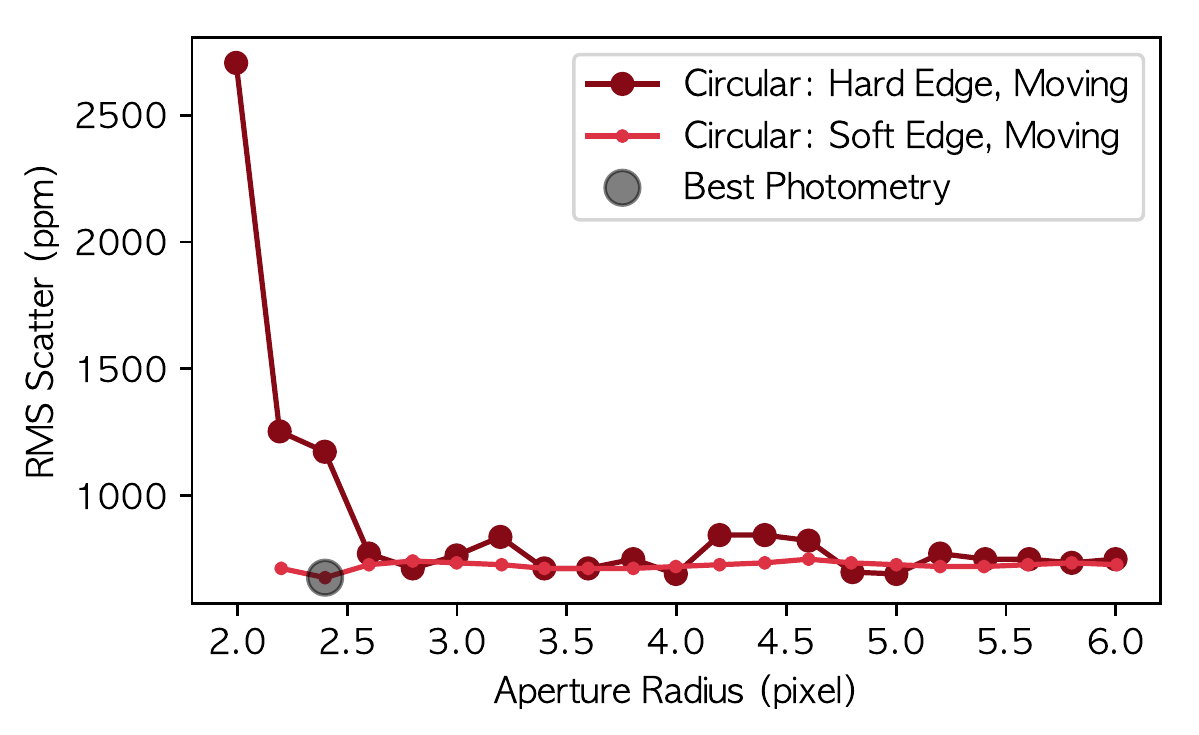}
\caption{The top and bottom panels show the root-mean-squared (RMS) scatter for various photometry schemes performed on 3.6~$\mu$m and 4.5~$\mu$m data, respectively. The schemes resulting in the smallest RMS scatter are hard-edge circular apertures with a radius of 3.0 pixels for the 3.6~$\mu$m observations and a soft-edge aperture of 2.4 pixel for the 4.5~$\mu$m observations; these are denoted by a gray circle in each panel. \label{fig:compare}}
\end{figure}



\begin{figure*}[!htbp]
\includegraphics[width=.495\linewidth, height=9.4cm]{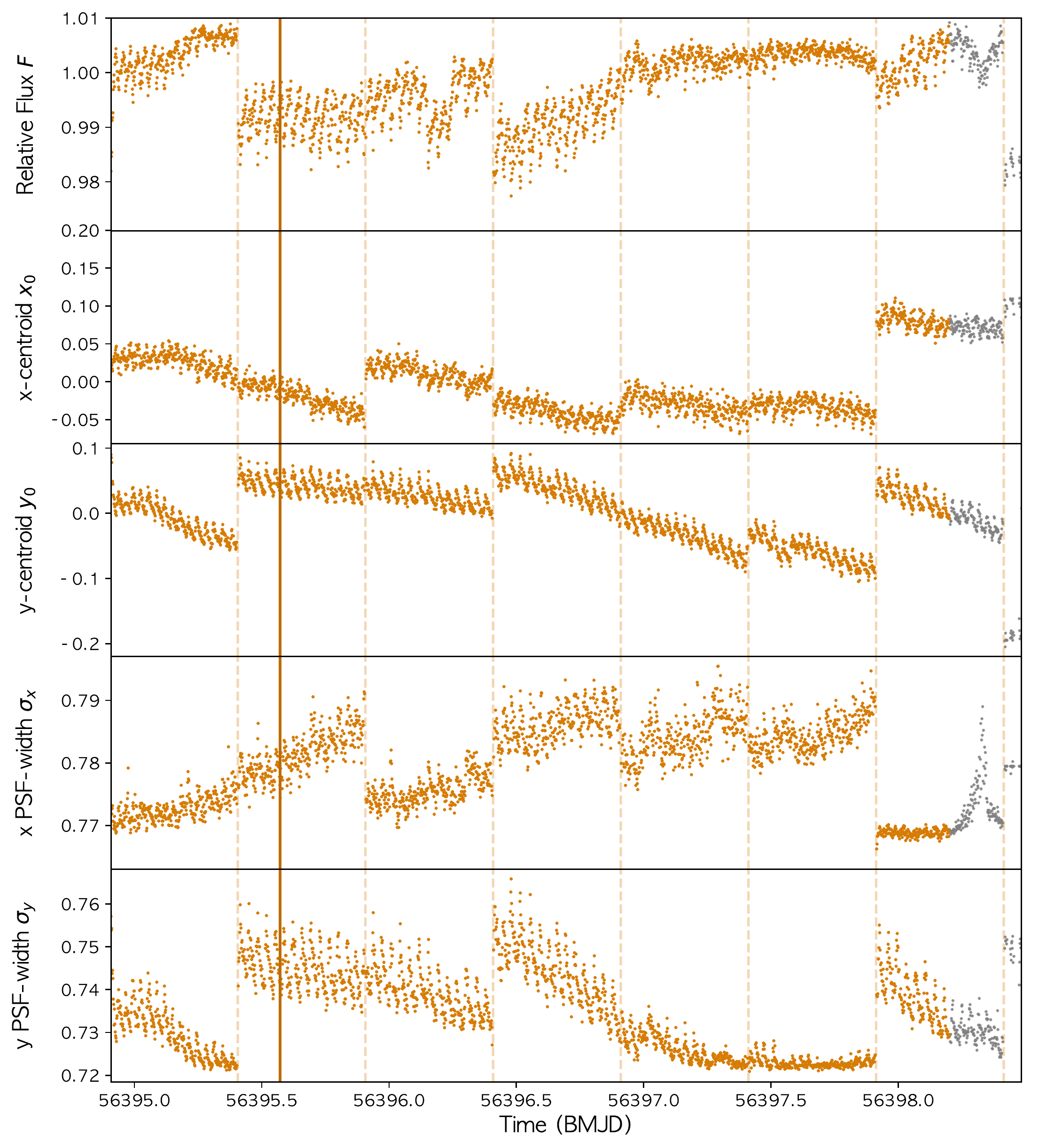}\includegraphics[width=.495\linewidth, height=9.4cm]{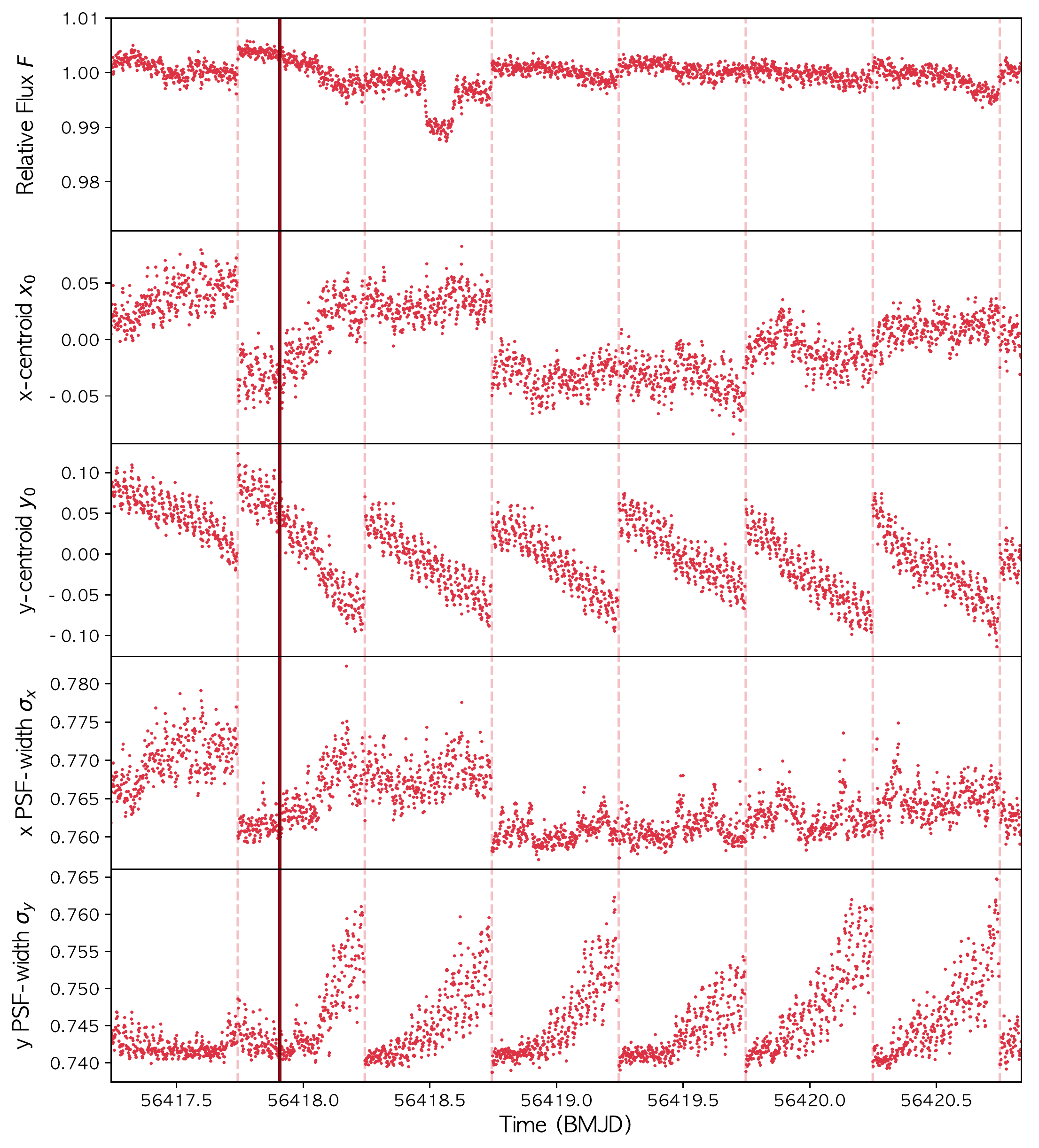}
\caption{Raw photometry  at  3.6~$\mu$m (left) and 4.5~$\mu$m (right). The top panel shows photometry using the preferred extraction scheme. The second and third panels show the $x$ and $y$ centroid coordinates, respectively. The fourth and fifth panels show PSF width along the $x$ and $y$ axes, respectively. The colored data points represent those used in the analysis while the gray points are discarded from the analysis. The vertical dark colored line represents the time of periastron passage. The vertical dashed lines represent the AOR breaks when the pointing of the spacecraft is readjusted. Note that the data were binned by 64-frame data cube for better visualization.  \label{fig:raw_photo}}
\end{figure*}

\begin{figure*}[!htbp]
\includegraphics[width=.495\linewidth]{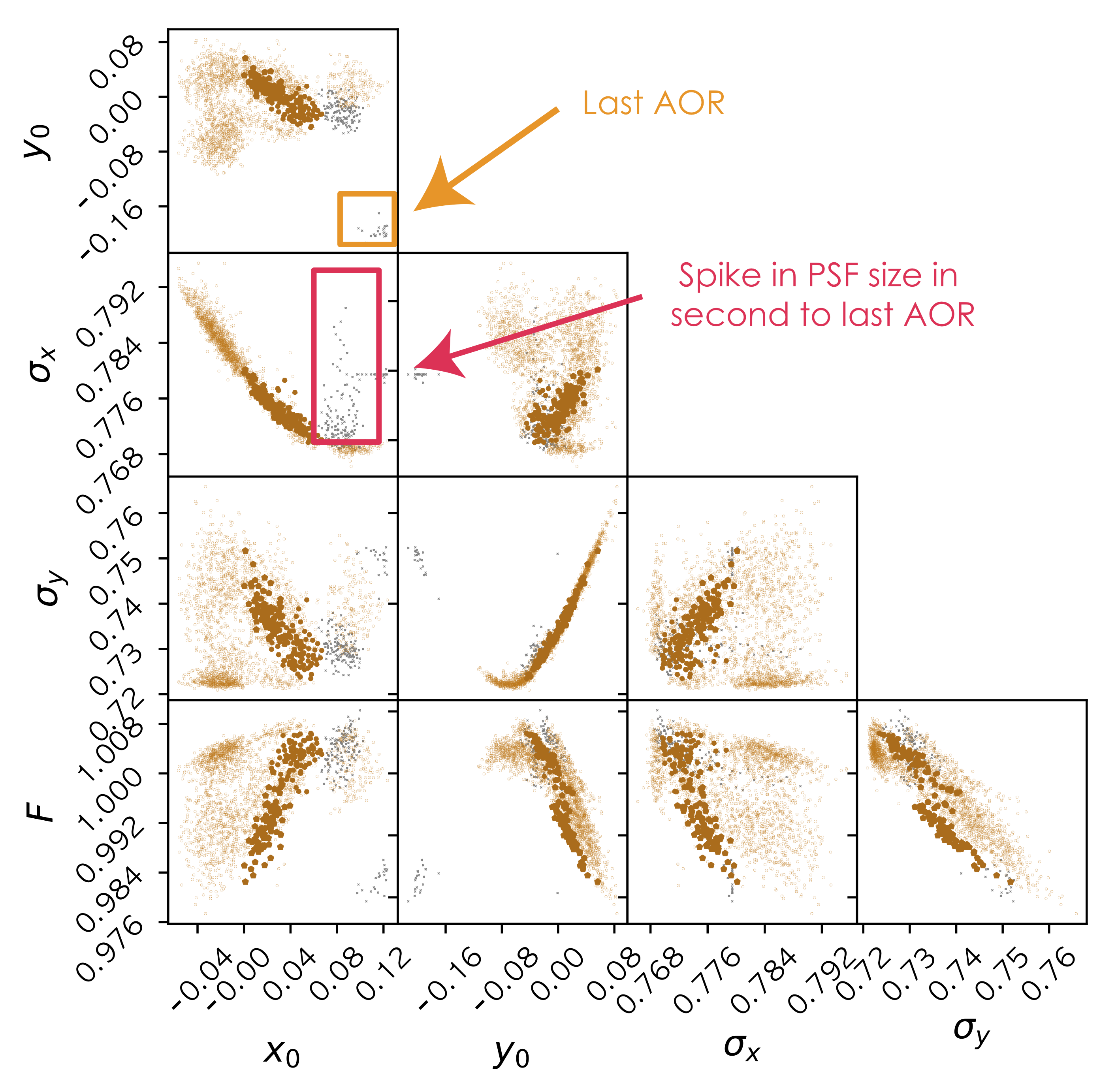}\includegraphics[width=.495\linewidth]{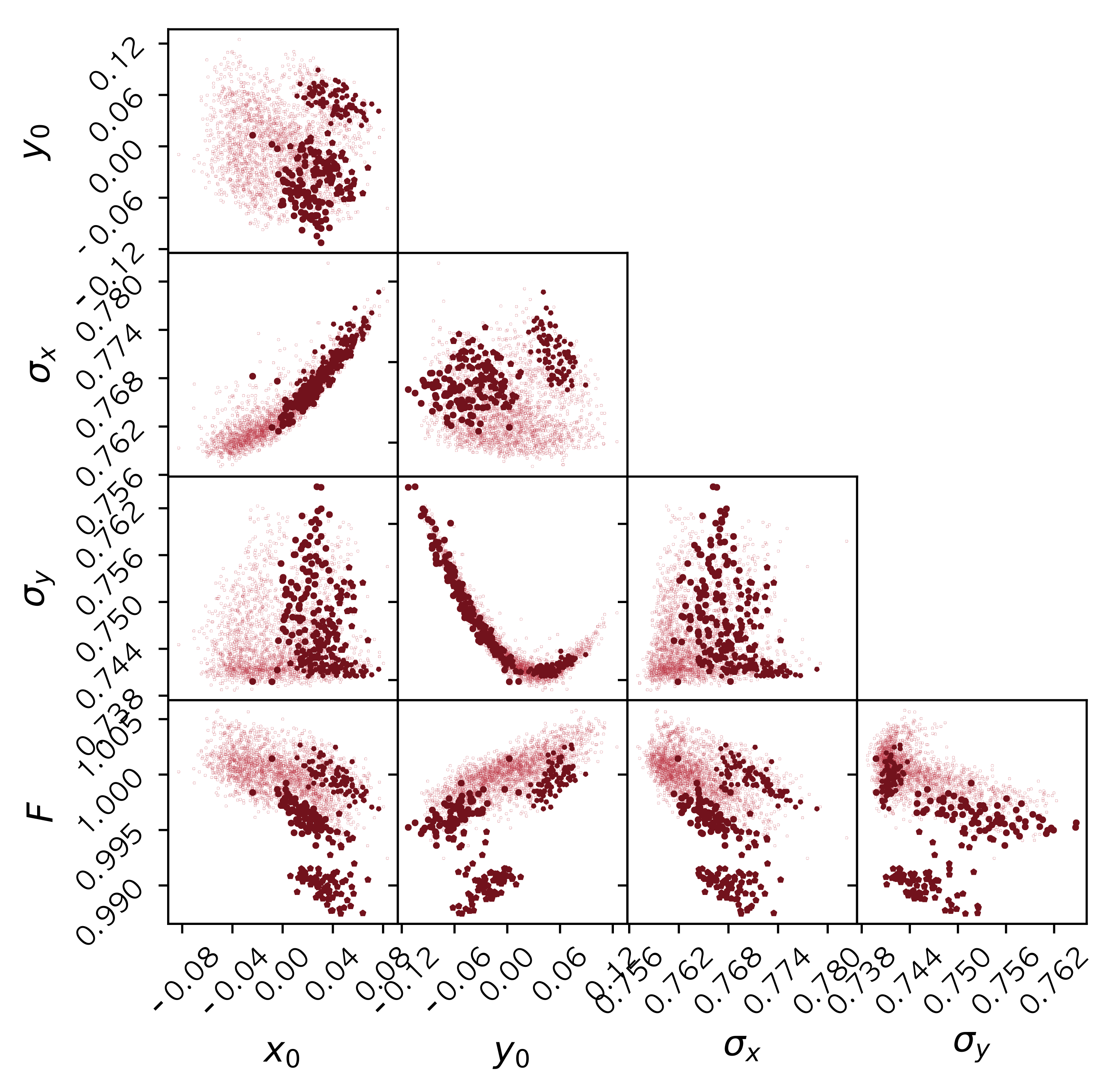}
\caption{PSF metrics for the 3.6~$\mu$m (left) and the 4.5~$\mu$m (right) observations. Each point represent the median of a datacube. The grey dots in the 3.6~$\mu$m plot represent data that were discarded from our analysis due to deviant PSF centroids or widths. The darker points represent the in-eclipse and in-transit points data.
\label{fig: psfmetric}}
\end{figure*}

\subsection{Data Reduction}

We use the Spitzer Science Center's basic calibrated data (BCD), which have been dark-subtracted, flat-fielded, linearized and flux calibrated using version S19.2.0 of the IRAC software pipeline.  \cite{2011ApJ...726...95D} first noted that the post-cryogenic \textit{Spitzer}/IRAC data collected in sub-array mode  exhibit frame-dependent background flux: they display a settling effect over the 64 frames as well as a sudden increase or decrease in background flux at the 58$^{\rm th}$ frame. \cite{2018NatAs...2..220D} found that data calibrated using the S19.2.0 pipeline exhibit frame-dependent flux modulation introduced during the sky dark subtraction stage. The 58$^{\rm th}$ frame error and the flux modulation are both fixed using correction image stacks for each AOR; these were provided by the IRAC team.

We use the Spitzer Phase Curve Analysis Pipeline\footnote{Details about how to install and use SPCA can be found at \texttt{https://spca.readthedocs.io}} \cite[SPCA;][]{2018NatAs...2..220D,2021MNRAS.504.3316B}, an open-source, modular, and automated pipeline to reduce and analyze our data. After correcting for frame-dependent systematics, we convert the pixel intensity from MJy/str to electron counts and obtain time stamps for each exposure using values from each \texttt{FITS} file.  We masked \textit{NaN} pixels and perform a pixel-by-pixel 4$\sigma$ sigma clip where the standard deviation, $\sigma$, for each pixel is determined along its respective datacube. We choose to mask rather than replace sigma-clipped pixels with average values to minimize the manipulation of the data. We then perform a pixel-level sigma-clipping by comparing each pixel with the pixel located at the same coordinate on all 64 frames of the same datacube and masking all 5$\sigma$ outliers. Frames containing a sigma-clipped pixel located in a 5$\times$5 pixel box centered on the central pixel of the target are discarded entirely.

We then evaluate the level of background flux for each frame by masking all the pixels within a 7$\times$7 pixel box centered on the target and measuring the median pixel intensity of the remaining unmasked pixels. We then perform background subtraction on each frame. Finally, we estimate the centroid coordinates ($x_0$, $y_0$) for each frame using the flux-weighted mean along the $x$ and $y$ axes and measure the Point-Spread-Function (PSF) width $(\sigma _{x_0}, \sigma _{y_0})$ along the $x$ and $y$ axes.



\subsection{Extracting the Photometry}

\subsubsection{Aperture Photometry}

We perform aperture photometry using soft-edge and hard edge circular apertures as defined in \cite{2021MNRAS.504.3316B} with radii from 1.25 to 7.25 pixels . We center the aperture on the PSF flux-weighted mean centroid of each frame. To determine the best photometric scheme, we calculate the root-mean-squared (RMS) scatter from a smoothed lightcurve by boxcar averaging with a length of 50. Figure~\ref{fig:compare} shows the resulting RMS for all our considered aperture choices; we select the scheme with the smallest RMS. We find that the best photometric schemes are hard-edge apertures with a radius of 3.0 and soft-edge apertures with a radius 2.4 pixels for the 3.6~$\mu$m and 4.5~$\mu$m channels, respectively. The raw photometry and PSF metrics are presented in Figure \ref{fig:raw_photo}.

\subsubsection{Pixel Level Decorrelation Photometry}
Pixel Level Decorrelation (PLD) models the systematics as a function of the fractional flux measured by each pixel within a stamp \citep{2015ApJ...805..132D}. SPCA's PLD photometry routine takes a $3 \times 3$ or $5 \times 5$~pixel stamp centered on the pixel position $(15,15)$.  The cleaning routine applied to each pixel lightcurve is described in \cite{2021MNRAS.504.3316B}.

\subsection{PSF Diagnostics}
As noted by \cite{2014A&A...572A..73L} and \cite{2021PSJ.....2....9C}, sharp fluctuations in the PSF width can alter the photometry. We search for anomalous PSF behavior by exploring the correlation between the PSF centroids and width shown in Figure \ref{fig: psfmetric}. Inspecting the PSF centroids for the 3.6~$\mu$m data, we find 2 distinct centroid clusters: a large cluster containing most of the data and a smaller cluster which corresponds to the centroid of the last AOR. Unfortunately, it is difficult to constrain the instrumental systematics of the smaller cluster due to the sparse data covering the area.  We therefore elect to discard the last AOR of 3.6~$\mu$m observations. 

As seen in Figure \ref{fig: psfmetric}, for both channels, the PSF width follows a parabolic function of centroid. However, with closer inspection of the PSF size plotted against the centroid along the x-direction of the 3.6~$\mu$m data, there is a significant deviation from the parabolic trend marked in gray. These deviant points corresponds to the greyed out points in the second to last AOR in Figure \ref{fig:raw_photo}. Note that this spike in PSF size coincides with a v-shape in the photometry and occurs during a secondary eclipse. The flux decrement is still seen using a larger aperture, which rules out the hypothesis that the inflated PSF causes some of the flux to fall outside the aperture. This detector systematic therefore remains unexplained but is presumably electronic in origin. We incorporate the PSF size into our detector sensitivity model as explained in Section \ref{sec:model}, but we were still unable to completely model out the instrumental signal. Upon analyzing the 2010 partial phase curve, we find that the spike feature led to an over-estimation of the eclipse depth. For this reason, we opt to discard these deviant points from our analysis.

Additionally, we elect to discard the first AOR from each phase curves containing 12 datacubes since the target was placed on a different pixel than the rest of the dataset for calibration purposes. After data reduction, a total of 2191 datacubes and 2388 datacubes are kept for analysis for the 3.6~$\mu$m and 4.5~$\mu$m channels, respectively. The products of our photometry extraction are shown in Figure~\ref{fig:raw_photo} and \ref{fig: psfmetric}. We then bin each dataset by datacube (64 frames) to reduce the computational cost of fitting the data with our many different decorrelation models.



\section{Model} \label{sec:model}

SPCA models the photometry as the product of the astrophysical model, $A(t)$, and the detector response, $D$: $F_{\rm total} = A(t) \times D$. Both models are evaluated simultaneously. 
We experiment with astrophysical models of varying complexity and with different parametric and non-parametric detector response models, as described below. This experiment results in a statistical analysis to determine which combination of astrophysical and detector model is preferred by the data.

\subsection{Astrophysical Model}

The stellar flux is assumed to be constant except during transit. The shape of the transit is modeled using \texttt{batman} \citep{2015PASP..127.1161K} with quadratic limb darkening. We modeled the astrophysical signal $A(t)$ as the sum of the stellar flux, $F_{\star}(t)$, and the planetary flux, $F_p(t)$:
\begin{equation}
A(t) = F_{\star}(t)+F_p(t).
\end{equation}
The planetary flux, $F_p = E(t) \times \Phi(t)$, is modeled as a sinusoidal phase variation multiplied by the secondary eclipse, $E(t)$, modeled assuming a uniform disk using \texttt{batman} \citep{2015PASP..127.1161K}. We did not account for the light travel time as it is only 45.02 seconds at superior conjunction and does not affect our analysis. The phase variation is modeled as a second order sinusoidal function, $\Phi$, and is expressed following \citep{2013ApJ...766...95L}:
\begin{equation}
\begin{split}
\Phi(\theta) = 1 & + A[\cos(\theta)-1] + B \sin(\theta) \\
& + C[\cos(2 \theta)-1] + D \sin(2 \theta),
\end{split}
\end{equation}
\noindent where $\theta = f + \omega + \pi /2$ is the orbital phase measured from mid-eclipse and $f$ and $\omega$ are the true anomaly and the argument of periastron, respectively. We also experiment with a first-order sinusoid with $C$ and $D$ set to 0 to determine the degree of complexity statistically preferred. The phase variation function is scaled such that it is 0 during secondary eclipse and $E(t)=F_p/F_*$ outside of occultation, where $F_p/F_*$ is the eclipse depth in terms of stellar flux. This parameterization allows for eclipse depth to be an explicit fit parameter. Despite its simple sinusoidal appearance, this parameterization captures the basic behaviour of more sophisticated simulations: rapid changes in flux near periastron when the planet's orbital phase and temperature both vary quickly, and slower flux evolution near apoastron.   

\begin{figure*}[!htbp]
\includegraphics[width=.495\linewidth]{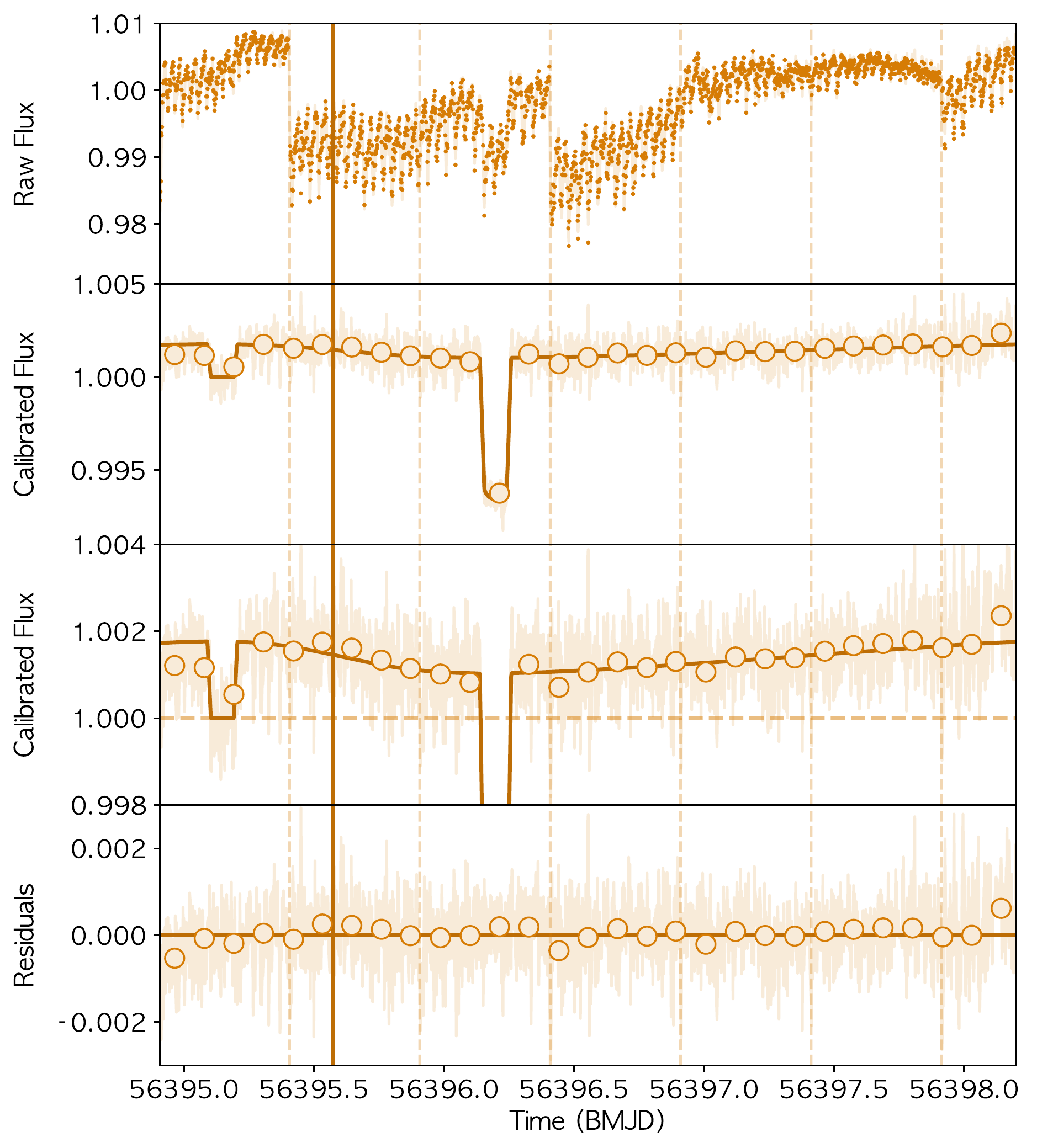}\includegraphics[width=.495\linewidth]{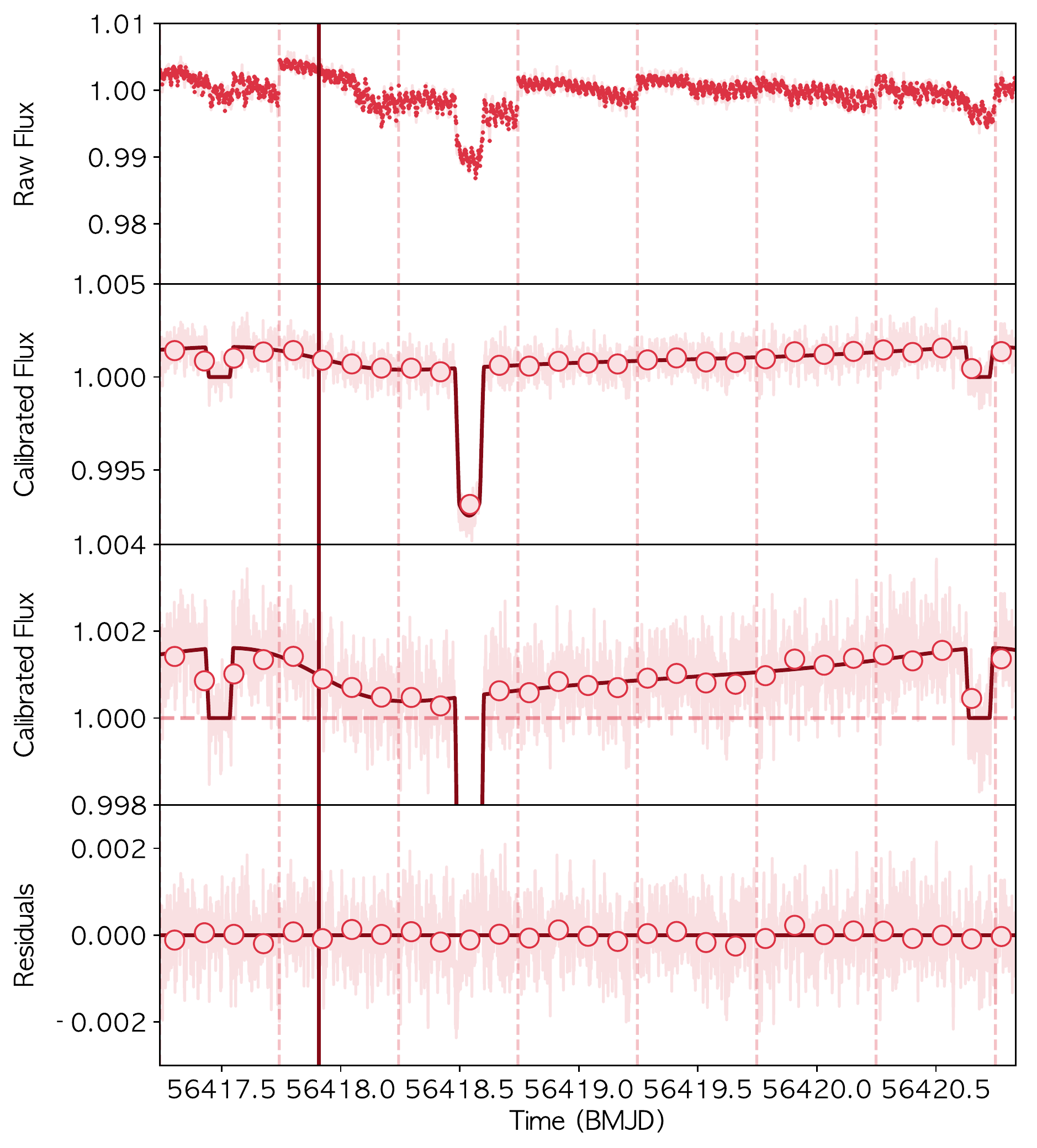}
\caption{The best-fit models to the 3.6~$\mu$m and 4.5~$\mu$m observations are presented in the left and right panels, respectively. The vertical dashed lines represent the AOR breaks and the dark vertical lines represent the time of periapse passage. The raw photometry is plotted in the top panels in pale yellow and red while the best-fit signal, $F_{\rm total}$, is shown in dark yellow and red. The corrected photometry is shown in the second panels in pale yellow and red and the best-fit astrophysical models are shown in darker colors. 
The third panels are a zoomed-in version of the second panels to more clearly show the phase variations. The last panels shows the residuals. Note that the 3.6~$\mu$m and 4.5~$\mu$m channels data were fitted independently. \label{fig:fit}}
\end{figure*}

\begin{figure*}[!htbp]
\includegraphics[width=.495\linewidth]{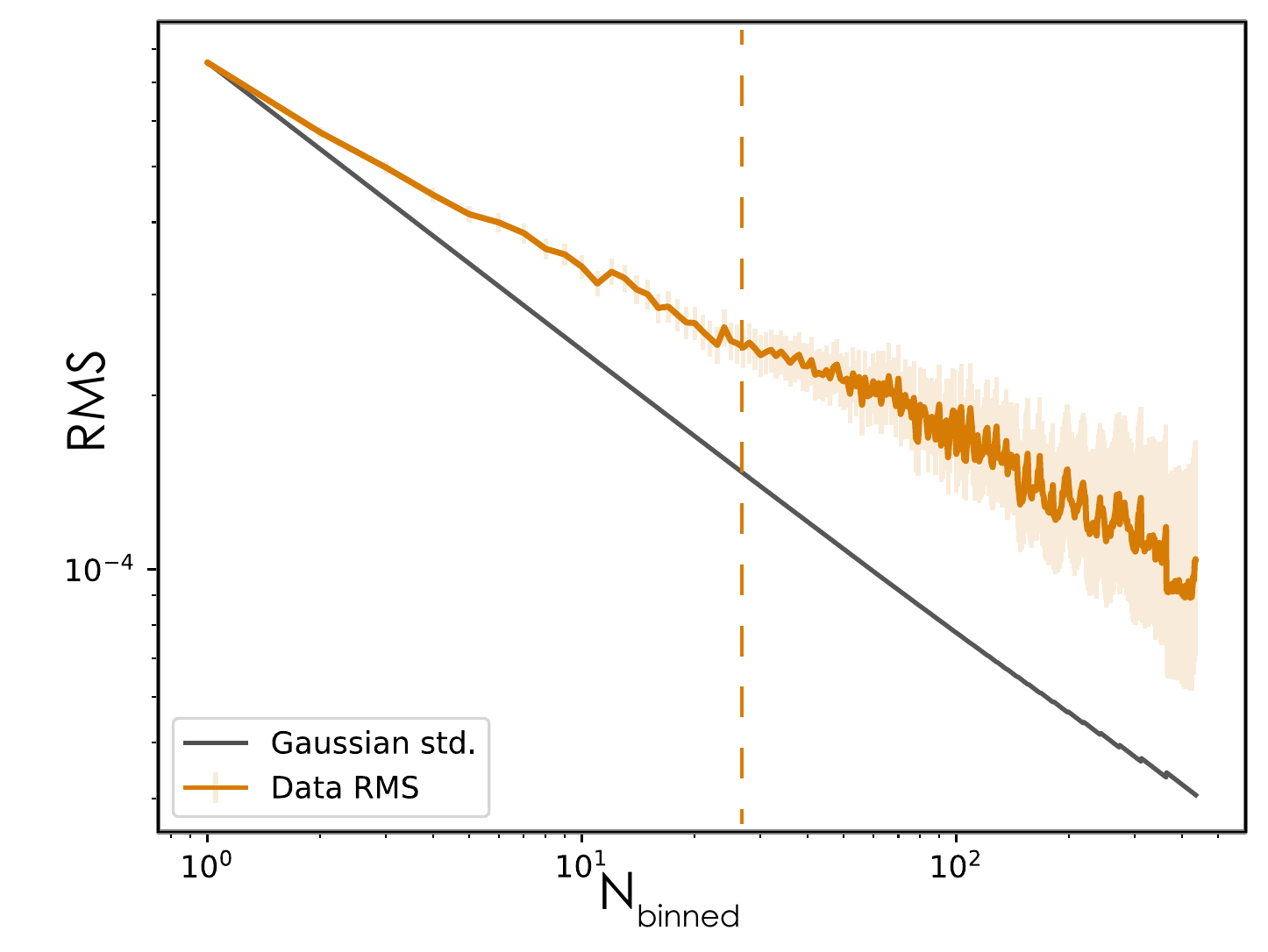}\includegraphics[width=.495\linewidth]{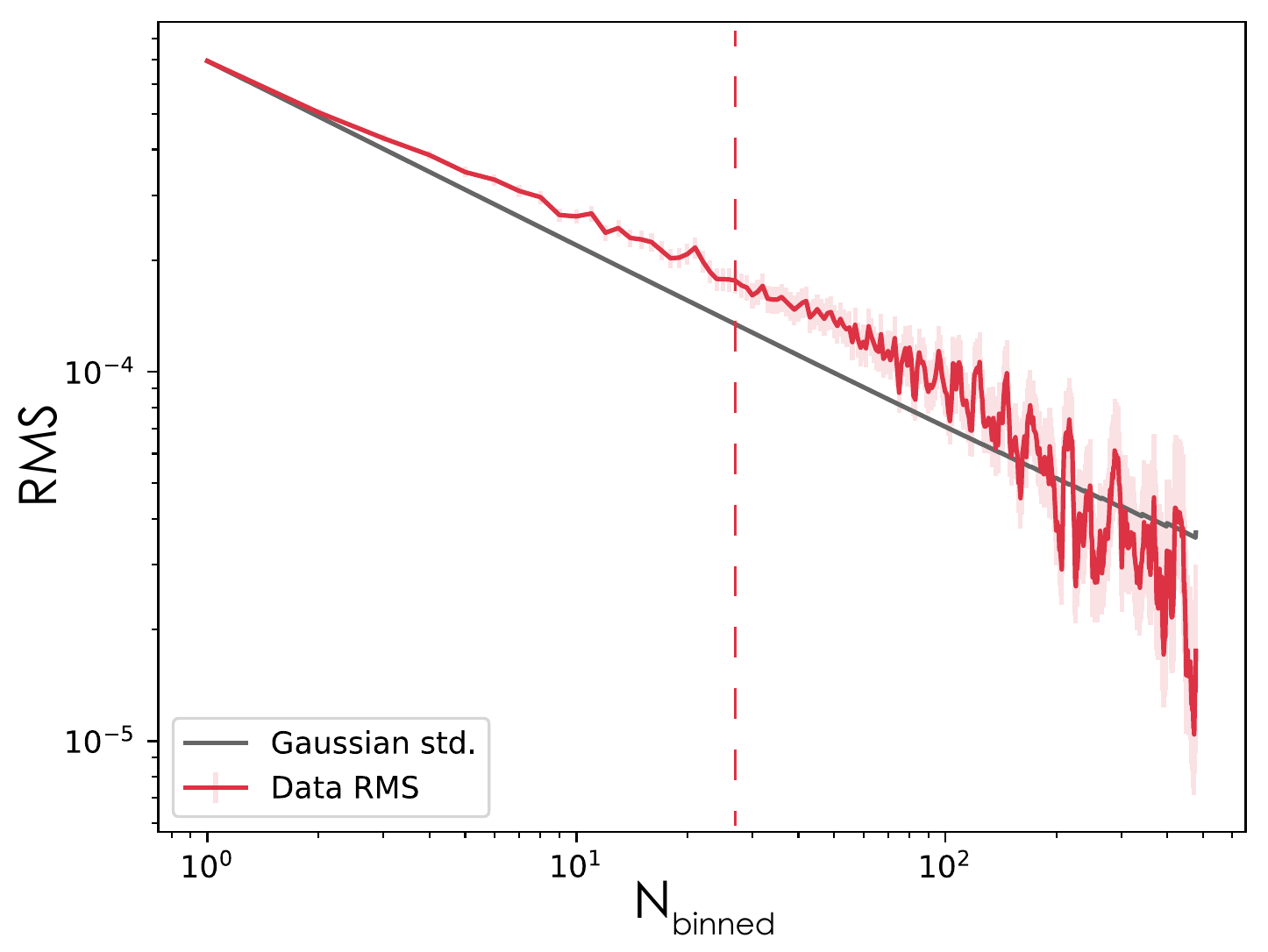}
\caption{Red-noise test for our best-fit models as a function of bin size for 3.6~$\mu$m (left) and 4.5~$\mu$m (right) fit. We binned the data into bins of different size, $N_{\rm binned}$, and computed the binned residuals RMS. The lighter shaded area is the uncertainty that the MC3 package computed \citep{2017AJ....153....3C}. The grey line represents the expected decrease in RMS if the residuals are purely white noise. We find that there is significant red noise in the residuals of the 3.6~$\mu$m fit, while the residuals of the 4.5~$\mu$m fit are less correlated. The vertical dashed line represents the number of bins contained in the duration of an eclipse depth. \label{fig:rednoise}}
\end{figure*}



\subsection{Detector Model}

The \textit{Spitzer}/IRAC 3.6~$\mu$m and 4.5~$\mu$m channels exhibit significant intrapixel sensitivity variations: for a given astrophysical flux, the electron count varies with the location and spread of the target's PSF on the detector \citep{2005ApJ...626..523C, 2014A&A...572A..73L}. Over the years, several decorrelation techniques have been developed to achieve an exquisite level of precision \citep[e.g.,][]{2016AJ....152...44I}. We tested several of these methods, namely 2D polynomials, BiLinear Interpolated Sub-pixel Sensitivity (BLISS) mapping, and Pixel Level Decorrelation (PLD). We briefly describe these methods below. 

\subsubsection{2D Polynomial Model}

The 2D polynomial model uses the PSF centroids as regressors and polynomial coefficients are fit parameters \citep{2005ApJ...626..523C, 2012ApJ...747...82C}. We experiment with second order to fifth order polynomials, including all cross terms.

\subsubsection{BLISS Mapping}

We also experiment with BLISS mapping, first proposed by \cite{2012ApJ...754..136S}. In summary, BLISS mapping is a data-driven iterative process to interpolate an intra-pixel sensitivity map of the central pixel, which will in turn be used to decorrelate the data. The area over which the PSF centroids are distributed is divided into subpixels, also called ``knots'', and each datum is associated with a knot. The data--astrophysical residuals are used to estimate the sensitivity of each knot and the sensitivity of each location of the detector is then estimated by bilinearly interpoating between the surrounding knots. We note that BLISS performs best with continuous uninterrupted \textit{Spitzer} observations. Since our observation scheme included more AOR breaks than most hot Jupiter phase curves, it may not be the best-suited dataset for this decorrelation method.

\subsubsection{Pixel-Level Decorrelation}
In contrast with BLISS mapping and polynomials, Pixel Level Decorrelation \citep{2015ApJ...805..132D, 2018AJ....156...99L} does not explicitly depend on the PSF centroids. Rather, this method uses the fractional flux measured by each pixel to model the detector systematics. In principle, astrophysical flux variations would change the intensities of all the pixels in the stamp that encompasses the target's PSF; however, they would not change the fractional fluxes of these pixels in the absence of detector systematics (e.g., spacecraft drift, thermal fluctuation). We experiment with $3 \times 3$ and $5 \times 5$ pixel stamps to explore the trade-off between capturing more stellar flux and more background flux. We test fits using a linear PLD and a second-order PLD that does not include cross-terms as they don't improve the quality of the fit \citep{2018AJ....155...83Z}.

\subsubsection{PSF Width}
Previous studies have shown that detector models including a function of the PSF width in $x$ and $y$ dramatically improve the photometric residuals \citep{2012ApJ...754...22K,2013ApJ...766...95L}. \cite{2020ApJ...888L..15M} and \cite{2021PSJ.....2....9C} experimented with linear,
quadratic, and cubic dependencies on the PRF width in both the $x$ and $y$ directions and find the linear model to be preferred. Hence, we include the function of PSF width, $D_{\rm PSFw}(\sigma _{x_0},\sigma _{x_0})$, as a multiplicative term to the detector models listed above:
\begin{equation}
\label{eq: psfw}
D_{\rm PSFw}(\sigma _{x_0},\sigma _{y_0}) = d_0 + d_1 \sigma _{x_0} + d_2  \sigma _{y_0}
\end{equation}

\noindent where $d_i$ are the fit parameters. We indeed find that including a PSF shape dependent model significantly reduces the red noise in the residuals when decorrelating data from both \textit{Spitzer}/IRAC channels. 

\subsubsection{Step Function}

For the 4.5 $\mu$m data, we could not get a good fit using any combination of the above detector models: the residuals showed a discontinuity between the first 2 AORs and the following 5 AORs. Often AOR discontinuities can be addressed by also fitting for a linear trend \citep[e.g.][]{2019MNRAS.489.1995B}, however, an unmodeled linear trend would exhibit a discontinuity at all the AOR breaks, which isn't the case here. To mitigate the problem, we added an additional detector sensitivity model as a multiplicative term:
\begin{equation}
\label{eq: step}
D_{step} = h_1[H(t-h_2)] + 1,
\end{equation}
where $H(t)$ is a Heaviside step function, $h_1$ is the amplitude of the DC offset, which we set as a fit parameter and $h_2$ is the time of the discontinuity which we fixed to the break between the 2$^{\rm nd}$ and 3$^{\rm rd}$ AOR. Our phase curve also includes 2 eclipses observations on either sides of the step to help constrain its magnitude.

\begin{deluxetable*}{l|ccc|cc}
\renewcommand{\arraystretch}{1} 
\tablecaption{Best-fit Model Parameter Estimates \label{tab:MCMCfit}\footnote{The uncertainties for the fitted parameters to the 3.6$~\mu$m data have been inflated by a factor of 1.72 to account for the correlated noise in the residuals (Fig.~\ref{fig:rednoise}).}}
\tablehead{
\colhead{Name} & \colhead{Symbol} & \colhead{Prior} & \colhead{Reference} & \colhead{3.6~$\mu$m} & \colhead{4.5~$\mu$m}
}
\startdata
\textbf{\textit{Fitted}}           & & & & &   \\
Time of transit (BMJD)    & $t_0$       & --                          & --                 &$56418.5434^{+0.0004} _{-0.0005}$ & $56418.5421\pm0.0002$ \\
Radius of planet          & $R_p/R_*$   & [0,1]                       & --                 &$0.0866^{+0.0014} _{-0.0012}$ & $0.0891\pm0.0008$ \\
Eclipse depth (ppm)        & $F_p/F_*$ & [0,1]                       & --                 &$1770\pm180$ &$1610\pm70$ \\
Limb Darkening coeff.     & $q_1$       & [0,1]                       & --                 &$0.05^{+0.06} _{-0.05}$ & $0.05^{+0.03} _{-0.02}$ \\
Limb Darkening coeff.     & $q_2$       & [0,1]                       & --                 &$0.2^{+0.6} _{-0.4}$ & $0.8^{+0.3} _{-0.4}$ \\
Phase var. coeff (order 1)& $A$         & --                          & --                 &$0.21\pm0.04$ & $0.34\pm0.02$ \\
Phase var. coeff (order 1)& $B$         & --                          & --                 &$-0.01\pm0.04$ & $-0.08^{+0.02} _{-0.03}$ \\
Phase var. coeff (order 2)& $C$         & --                          & --                 & -- &$0.05\pm0.02$ \\
Phase var. coeff (order 2)& $D$         & --                          & --                 & -- &$0.07^{+0.01} _{-0.02}$ \\
\hline
\textbf{\textit{Fixed}}   &             &                             &                    &  &  \\
Period (days)             & $P$         & 3.19153285 $\pm$ 0.000000058    & Wong et al. (2014) & -- & -- \\
Semi-major axis           & $a/R*$      &  7.052 $^{+0.076} _{-0.097}$    & Wong et al. (2014) & -- & -- \\
Inclination               & $i$         &  84.11  $\pm$ 0.16              & Wong et al. (2014) & -- & -- \\
Eccentricity              & $e$         &  0.2769 $^{+0.0017} _{-0.0016}$  & Wong et al. (2014) & -- & -- \\
Longitude of periapse     & $\omega$    &  347.2 $^{+1.7} _{-1.6}$  & Wong et al. (2014) & -- & -- \\
\hline
\textbf{\textit{Derived}}         & & & & &   \\
Periastron passage after transit (days) & -- & 2.5645 & -- & -- & -- \\
Phase curve peak after transit (days) & -- & -- & -- & 2.11 $\pm$ 0.09 & 2.17 $\pm$ 0.03 \\
Phase curve trough after transit (days) & -- & -- & -- & -0.04 $\pm$ 0.10 & -0.27 $\pm$ 0.04 \\
Phase curve peak after eclipse (hrs)    & -- & -- & -- & -0.53 $\pm$ 1.98 & 0.95 $\pm$ 0.82\\
\enddata
\end{deluxetable*}


\section{Parameter Estimation and Model Comparison} \label{sec:Param}

We use the Affine Invariant Markov Chain Monte Carlo (MCMC) Ensemble Sampler from the \texttt{emcee} package to estimate the parameters and their respective uncertainties \citep{2013PASP..125..306F}. We elect to fix the orbital period, $P$, the semi-major axis, $a$, the inclination, $i$, the eccentricity, $e$, and the argument of periastron, $\omega$, to the values reported by \cite{2014ApJ...794..134W}. We opt to fix the values rather than to impose a Gaussian prior on these parameters to significantly improve the analysis runtime. Although using fixed values instead of distributions can lead to an underestimation of the other model parameter uncertainties, the published uncertainties on the fixed parameters represent less than 0.1\% of the reported value and therefore their contribution is negligible. 

We initialize the astrophysical parameters to be the values reported by \cite{2014ApJ...794..134W}. We begin an initial stage of parameter optimization as described in \cite{2021MNRAS.504.3316B}. We require that transit depths and eclipse depths be between zero and unity. We use the parametrization of \cite{2013MNRAS.435.2152K} for the limb-darkening coefficients to ensure that our walkers only explore physically plausible solutions with uniform uninformative sampling. 

By default, our pipeline includes a prior rejecting all models with phase variation coefficient that yield negative phase curves \citep{2017ApJ...849L...5K}. Since an eccentric planet like XO-3b is subject to eccentricity seasons and is expected to have a time-variable atmosphere, we cannot map the planet following \cite{2008ApJ...678L.129C} and hence cannot apply the more stringent constraint that the implied planetary map is non-negative \citep{2017ApJ...849L...5K,2019NatAs...3.1092K}. Nonetheless, our ability to fit the phase variations with an energy balance model in \S~\ref{sec:ebm} demonstrates that they do not require regions with negative flux. In any case, due to the deep eclipse depths, the fraction of rejected phase curves is less than 0.0001\%.

We make the photometric uncertainty, $\sigma _F$, a fitted parameter and use \texttt{emcee} to estimate the set of parameters that maximizes the log-likelihood:
\begin{equation}
\ln(L) = -\frac{1}{2}\chi ^2 - N_{\rm dat}\ln \sigma _F - \frac{N_{\rm dat}}{2} \ln (2\pi),
\end{equation}
where $N_{\rm dat}$ is the number of data. The badness-of-fit is defined as
\begin{equation}
\chi ^2 = \frac{\sum _i ^{N_{\rm dat}} [F_i - F_{i,\rm model}]^2}{\sigma _F ^2},
\end{equation}
where $F_i$ are brightness measurements and $F_{i,\rm model}$ are the predicted brightness from the astrophysical model described in section \ref{sec:model}. 

We initialize 300 MCMC walkers as a Gaussian ball distributed tightly around our initial guess. We perform an initial burn-in to let the walkers explore a wide region in parameter space during which each walker performs 5000 steps. We then perform a 1000 steps production run while making sure we meet our convergence criteria: 1) the log-likelihood of the best walker does not change over last 1000 steps of the MCMC chain and 2) the distribution of walkers is constant over the last 1000 steps along each parameter. We then obtain a posterior distribution and estimate the 1$\sigma$ confidence region as the 16$^{th}$ to 84$^{th}$ percentile of the posterior distribution of each parameter using all the walkers over the last 1000 MCMC steps.



\subsection{Model Comparison}

\begin{figure*}
\includegraphics[width=\linewidth]{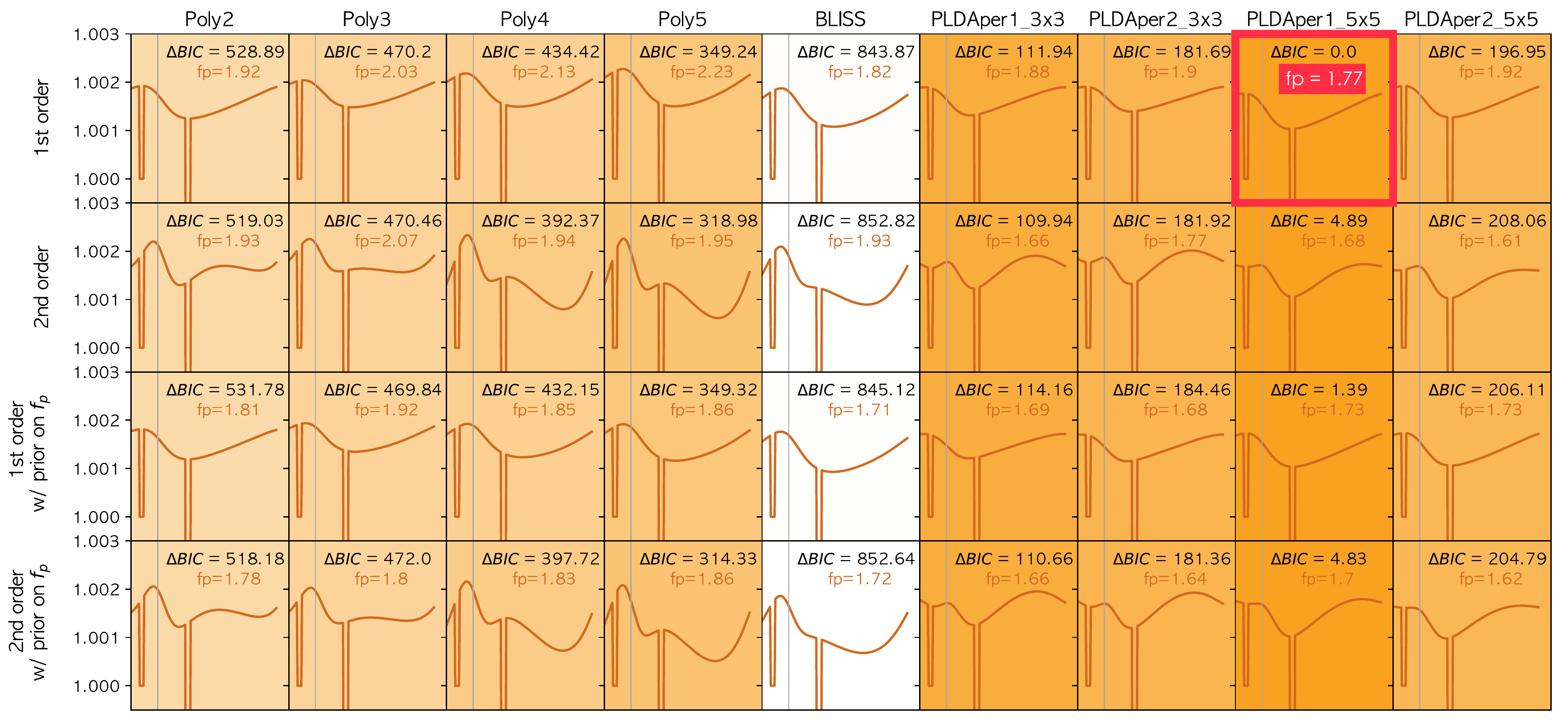}\\
\includegraphics[width=\linewidth]{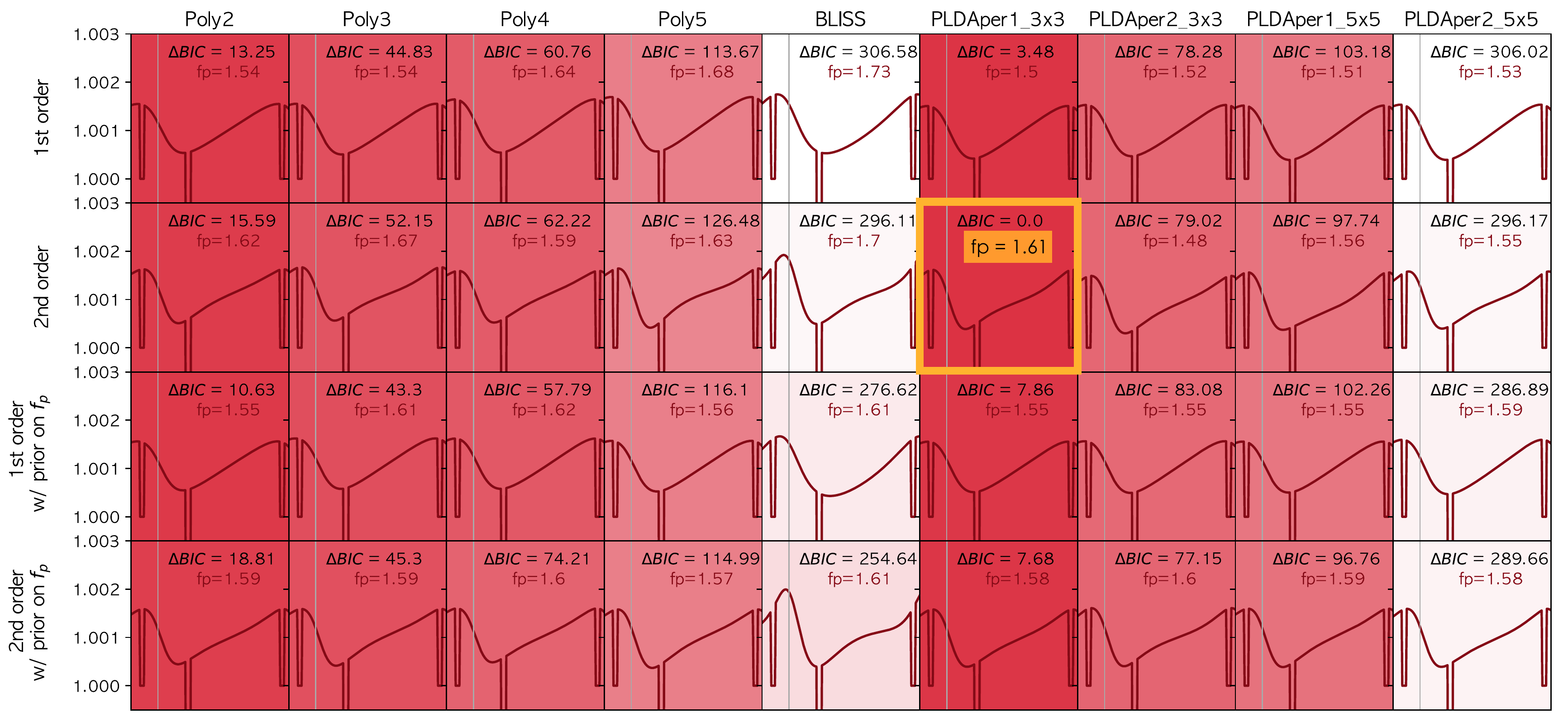}
\caption{Model comparisons for Spitzer $3.6~\mu$m (\textit{top}) and $4.5~\mu m$ (\textit{bottom}) phase variations of XO-3b. The best-fit astrophysical model to the \textit{Spitzer} data obtained using different detector models and phase variation models are shown here. Each column indicates the detector model used. The rows indicate the phase variation model used and whether or not a prior was imposed on the secondary eclipse depth. The $\Delta$BIC from the preferred solution of each fit are indicated in each box and the opacity of the background of each box reflects fit preference (darker is better). The eclipse depth for each fit is denoted by $f_p$, in parts-per-mille. \textit{Top}: The fit using a first-order PLD with a $5 \times 5$ stamp using a first-order sinusoidal model is preferred. In principle, the shape of the best-fit phase curve is dependent on the models and priors chosen. Fortunately, the shape of the first-order phase variation model is robust against the choice of detector model and prior on the secondary eclipse depth. \textit{Bottom:} The preferred fit uses a first-order PLD with a $3 \times 3$ stamp and a second-order sinusoidal model. Note that the shape of the 4.5 $\mu$m phase curves is robust against prior and model choices, but the 3.6 $\mu$m phase curve shape is not. \label{fig:fit_grid}}
\end{figure*}

As mentioned previously, we experiment with various astrophysical and detector response models. Generally, when the number of fit parameters increase, the fit to the data also improves since the model becomes more flexible. Instead of comparing the badness-of-fit or log-likelihood of the best-fit obtained by each model, we therefore compare the Bayesian Information Criterion \citep[BIC;][]{schwarz1978estimating} of the different models:
\begin{equation}
BIC = N_{\rm par}\ln N_{\rm dat} - 2 \ln L,
\end{equation}
where $N_{\rm par}$ is the number of fit parameters. By definition, a smaller BIC is preferred. A comprehensive model comparison can be found in Figure \ref{fig:fit_grid} which shows the shape of the astrophysical phase variation for each model fit and their relative BIC. In principle, the more fit parameters a model has, the more flexible it is and the better goodness-of-fit it achieves. The BIC allows us to justify or rule out having a more complex model with more fit parameters. We note that BLISS does not have any explicit fit parameters, instead we use the number of BLISS knots as number of parameters to estimate its BIC. This could explain why BLISS seems to exhibits systematically worst BIC than the other decorrelation methods (Figure \ref{fig:compare}), however, we note that the BLISS phase curves also have systematically different amplitudes and shapes than that of the phase curves retrieved using different decorrelation methods.


\section{Results}\label{sec:results}


\subsection{SPCA Fits}

As shown in Figure \ref{fig:fit_grid}, for our 3.6 $\mu$m phase observations, the first-order PLD model using a 5$\times$5 pixels stamp and a first-order sinusoidal phase curve was the preferred model. For our 4.5 $\mu$m observations, the preferred solution is a first-order PLD model using a 3$\times$3 pixels stamp and a second-order sinusoidal phase curve. Both preferred fits are shown in detail in Figure \ref{fig:fit}. We note that the 3.6~$\mu$m fit leaves noisier residuals than the 4.5~$\mu$m. Figure \ref{fig:rednoise} shows a red-noise test performed on the residuals of the 3.6~$\mu$m and 4.5~$\mu$m fits. The black line on both plots represent the expected decrease in root-mean-squared (RMS) scatter when uncorrelated data are binned. While the 4.5 $\mu$m RMS is in good agreement with this line, the larger than expected RMS of the 3.6 $\mu$m channel is indicative of leftover detector or astrophysical variations that our model could not fit. 

The best-fit astrophysical parameters for the \textit{Spitzer} observations are presented in Table \ref{tab:MCMCfit}. We find a ratio of planet to star radius of $R_p/R_*=0.0866^{+0.0014} _{-0.0012}$ and $0.0891\pm0.0008$ with the 3.6~$\mu$m and 4.5~$\mu$m observations, respectively. We measure secondary eclipse depths of $1770 \pm 180$ ppm and $1610 \pm 70$ ppm at $3.6~\mu$m and $4.5~\mu$m, respectively. The 4.5~$\mu$m eclipse depth is within 1$\sigma$ of that reported by \cite{2015ApJ...811..122W} and \cite{2016AJ....152...44I}. While both phase curves were analyzed independently of each other, their shapes are similar (Figure \ref{fig:temperature}). The 3.6 $\mu$m and 4.5 $\mu$m phase curves peak $2.11 \pm 0.09$ and $2.17 \pm 0.03$ days after transit, respectively. The minimum of the phase curves occur $0.04 \pm 0.10$ and $0.27 \pm 0.04$ days before transit. We also note that the 3.6 $\mu$m and 4.5 $\mu$m best-fit transit time differs by $\Delta t_0 = 0.0013 \pm 0.0005$ days. This less than 3 $\sigma$ discrepancy is about the temporal resolution of our phase curve, hence the difference in the 3.6 and 4.5 $\mu$m $t_0$ isn't meaningful.

\subsection{Energy Balance Model}\label{sec:ebm}

\begin{figure*}
\includegraphics[width=\linewidth]{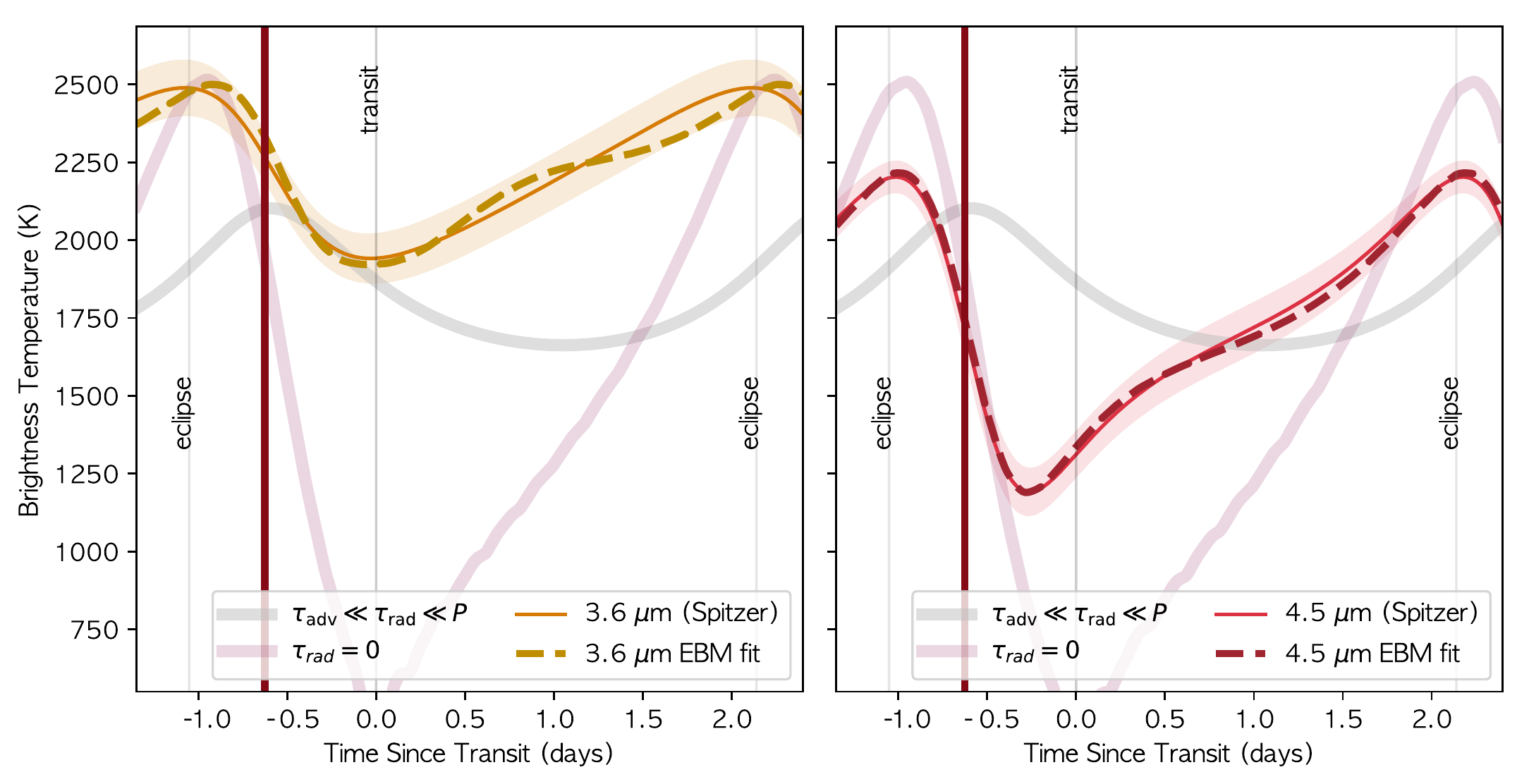}
\caption{The yellow (\textit{left}) and red (\textit{right}) line and swath represent the brightness temperature and uncertainties of XO-3b based on the \textit{Spitzer} 3.6~$\mu$m and 4.5~$\mu$m observations, respectively. The dark yellow and red dashed lines represent the energy balance model fits to the temperature curves. The light grey lines represent the time of secondary eclipse and the dark grey line indicates the transit time. The vertical brown line represents the periastron passage. In both panels, the grey transparent curve represent the limiting case with a short advective timescale $\tau _{\rm adv} \ll \tau _{\rm rad} \ll P$ and the pink transparent line represents the limiting case of a short radiative timescale, $\tau _{\rm rad}=0$.\label{fig:temperature}}
\end{figure*}

\begin{deluxetable*}{l|cc|cc}
\renewcommand{\arraystretch}{.78} 
\tablecaption{\textbf{Energy Balance Model Fit} \label{tab:EBM_fit_largep0}}
\tablehead{
\colhead{Parameter} & \colhead{Prior} & \colhead{Reference} & \colhead{3.6~$\mu$m} & \colhead{4.5~$\mu$m}
}
\startdata
\textbf{\textit{Fitted}} & & & & \\
$P_0$ (bar) & [$10^{-3}$, 20] & -- & $0.15 ^{+0.21} _{-0.11}$ & $2.40 ^{+0.92} _{-0.16}$ \\
Equatorial Wind Speed, $V_{\rm{wind}}$, (km/s) &  -- & -- & $12.1 ^{+3.0} _{-4.4}$ & $3.13 ^{+0.26} _{-0.83}$\\
Internal Energy, $E_{\rm{int}}$, ($W/m^2$) & [0, $+ \infty$] & -- & $(7.21 ^{+0.11} _{-0.17}) \times 10^5$ & $(1.6  ^{+11.8} _{-27.0}) \times 10^2$\\ 
Albedo, $A_B$ & [0,1] & -- & $0.0$ & $0.106 ^{+0.008} _{-0.106}$ \\
\textbf{\textit{Fixed}} & & & & \\
Period, $P$, (days)        & 3.19153285 $\pm$ 0.000000058    & Wong et al. (2014) & -- & -- \\
Eccentricity, $e$          & 0.2769 $^{+0.0017} _{-0.0016}$  & Wong et al. (2014) & -- & -- \\
Semi-major axis, $a$, (AU) &  0.04589 $^{+0.0063} _{-0.0049}$       & Wong et al. (2014) & -- & -- \\
Planet Radius, $R_p$, ($R_{\rm{Jup}}$) & 1.295 $\pm$ 0.066    & this work & -- & -- \\
Planet Mass, $M_p$, ($M_{\rm{Jup}}$)   & 11.79 $\pm$ 0.98     & this work & -- & -- \\
Inclination, $i$, (deg)    &  84.11  $\pm$ 0.16          & Wong et al. (2014) & -- & -- \\
Longitude of periapse, $\omega$  &  347.2 $^{+1.7} _{-1.6}$    & Wong et al. (2014) & -- & -- \\
Pseudo-Synchronous Rotation Period, $P_{\rm{ps}}$ (days) & 2.170 & Hut (1981) prescription & -- & -- \\
Stellar Effective Temperature, $T_{\rm{eff}}$, (K) & $6759 \pm 79$ & Soubiran et al. 2020  & -- & -- \\
Stellar Radius, $R_{*}$, ($R_{\odot}$) & $1.407 \pm 0.038$ & this work & -- & -- \\
Stellar Mass, $M_{*}$, ($M_{\odot}$) & 1.21 $\pm$ 0.15 & this work & -- & -- \\
\textbf{\textit{Derived}} & & & & \\
Average Radiative Timescale, $\tau _{\rm{rad, ave}}$, (hrs) & -- & -- & $\sim$ 6
& $\sim$ 30 \\
Advective Timescale, $\tau _{adv}$, (hrs) & -- & -- & $\sim$ 0.1 
& $\sim$ 1 \\
\enddata
\end{deluxetable*}

To extract radiative and advective properties of the atmosphere of XO-3b, we use the \texttt{Bell\_EBM} semi-analytical energy balance model (EBM) to fit our detrended phase variations \citep{2018ApJ...857L..20B}. As the atmospheric temperature does not exceed 2500~K, we did not include the effect of recombination and dissociation of hydrogen. The model treats the advection of atmospheric gas via solid body rotation at angular velocity $\omega_{\rm wind}$, assumes a uniform Bond albedo, $A_B$, and a uniform atmospheric pressure, $P_0$, at the bottom of the mixed layer (the part of the atmosphere that responds to diurnal and seasonal forcing). These quantities are treated as fit parameters and remain constant throughout the orbit. Depending on the efficiency of turbulent mixing \citep{2010ApJ...721.1113Y, 2018ApJ...854....8B} and varying barotropic large-scale flows vertical thickness, $P_0$ could be deeper in the atmosphere than the pressure at which incoming optical light is absorbed.

Given the significant correlation in the 3.6~$\mu$m residuals and that different wavelengths probe different photospheres \citep{2017ApJ...851L..26D}, we opt to evaluate each phase curve separately. The energy balance model is fit to the phase curves using the MCMC package \texttt{emcee} \citep{2013PASP..125..306F}. Again, we decide to fix the orbital period, $P$, the semi-major axis, $a$, the inclination, $i$, the eccentricity, $e$, and the argument of periastron, $\omega$, with values reported in Table \ref{tab:EBM_fit_largep0}. We use the stellar effective temperature from \textit{Gaia} DR2 \citep{2018A&A...616A...1G} as well as the updated radius and mass of XO-3 based on \textit{Gaia} DR2 parallaxes \citep[Stassun et al. in prep]{2017AJ....153..136S}. 

The \texttt{Bell\_EBM} is agnostic about the underlying rotation of XO-3b, since its atmosphere is not expected to remain stationary with respect to the deeper regions. Nonetheless, in order to convert the inertial-frame atmospheric angular frequency, $\omega_{\rm atm}$, to a zonal wind velocity, we must assume a rotational frequency for the interior of the planet.   The interiors of short-period eccentric planets are expected to be pseudo-synchronously rotating. Roughly speaking, this means that the planet is momentarily tidally locked near periapse passage, when the tidal forces are strongest.  We adopt the prescription of \cite{1981A&A....99..126H} for the pseudo-synchronous rotation frequency,  $\omega _{ps} \simeq 0.8 \omega _{\rm max}$, where the maximum orbital angular velocity at periastron is \citep{2011ApJ...726...82C}:
\begin{equation}
    \omega _{\rm max} = \frac{2\pi}{P} \frac{(1+e)^{1/2}}{(1-e)^{3/2}}.
\end{equation}
The equatorial wind velocity is therefore \begin{equation}
    v_{\rm wind} \mathbf{\simeq} (\omega_{\rm atm} - \omega_{ps})R_p.
\end{equation}

We transform the \textit{Spitzer} phase curve into an orbital apparent brightness temperature profile and fit it with our energy balance model. Physically motivated uniform priors are imposed to the fit parameters (see Table \ref{tab:EBM_fit_largep0}). Our initial attempts to fit the 3.6 $\mu$m \textit{Spitzer} phase-dependent temperatures with the \texttt{Bell\_EBM} could not reproduce the high brightness temperatures at all orbital phases. Due to the eccentricity of XO-3b's orbit, the planet is expected to experience tidal heating. We therefore add an internal energy source flux term, $E_{\rm{int}}$, to equation (1) of \cite{2018ApJ...857L..20B} and fit for this extra parameter. We experiment with and without this term and find that models allowing for an internal energy source are preferred. The best-fit EBM models to each channel are presented in Table \ref{tab:EBM_fit_largep0} and Figure \ref{fig:temperature}. For comparison, we also show the temperature curve of a limiting case with a very short advective timescale $\tau _{\rm adv} \ll \tau _{\rm rad} \ll P$, such that the incident energy is uniformly redistributed instantaneously (the advection-dominated phase curve). We also show the special case where $\tau _{\rm rad}=0$, i.e., the radiation-dominated phase curve. Neither of these limiting cases account for the presence of an internal energy source.

\subsubsection{\SI{4.5}{\um} EBM Fit}
The energy balance model is able to reproduce the timing and amplitude of the 4.5 $\mu$m phase curve peak and the minimum is only 1$\sigma$ from the \textit{Spitzer} observations. 
The best-fit model has $v_{\rm wind}$ of $3.13 ^{+0.26} _{-0.83}$ km/s: approximately the speed of sound. The model suggests that the mixed layer extends down to $P_0 = 2.40 ^{+0.09} _{-0.16}$ bar and a Bond albedo $A_b=0.106 ^{+0.008} _{-0.106}$. From these, we estimate an advective timescale of $\sim 1$ hour and an average radiative timescale of $\sim 30$ hours. Unlike the 3.6 $\mu$m phase curve, we do not need internal heating to explain the 4.5 $\mu$m phase curve.


\subsection{Spitzer/IRAC \SI{3.6}{\um}: A Cautionary Tale}
\subsubsection{Strong Detector Systematics}
\textit{Spitzer}/IRAC's 3.6 $\mu$m channel is known to be less stable than the 4.5 $\mu$m channel: stronger detector systematics generally plague the 3.6 $\mu$m observations \citep[e.g.][]{2018AJ....155...83Z}. As discussed in Section 2.3, the 3.6 $\mu$m observations exhibit a sharp PSF fluctuation coinciding with one of the secondary eclipses. We experiment with and without excluding the anomalous observations and find an eclipse depth $2130 \pm 110$ ppm when the aberrant observations are included and $1770 \pm 100$ ppm when discarded. The deeper eclipse depth estimate is likely a result of the large decrease in flux caused by the sharp PSF width fluctuation, hence we elect to omit the anomalous portion of the observations for the analysis. Consequently, without a reliable second eclipse, it is significantly more difficult to distinguish astrophysical trends from detector systematics. Furthermore, Figure \ref{fig:rednoise} indicates that the residuals from our best-fit model to XO-3b's 3.6 $\mu$m phase curve are significantly correlated; on the eclipse duration timescale the 3.6~$\mu$m residual RMS is 1.72 times larger than expected if the residuals were uncorrelated. Hence, we elect to inflate our SPCA fit uncertainties by 1.72 for the 3.6~$\mu$m fit and the 3.6 $\mu$m observations should be interpreted with caution. 

\begin{figure}[!htpb]
\includegraphics[width=\linewidth]{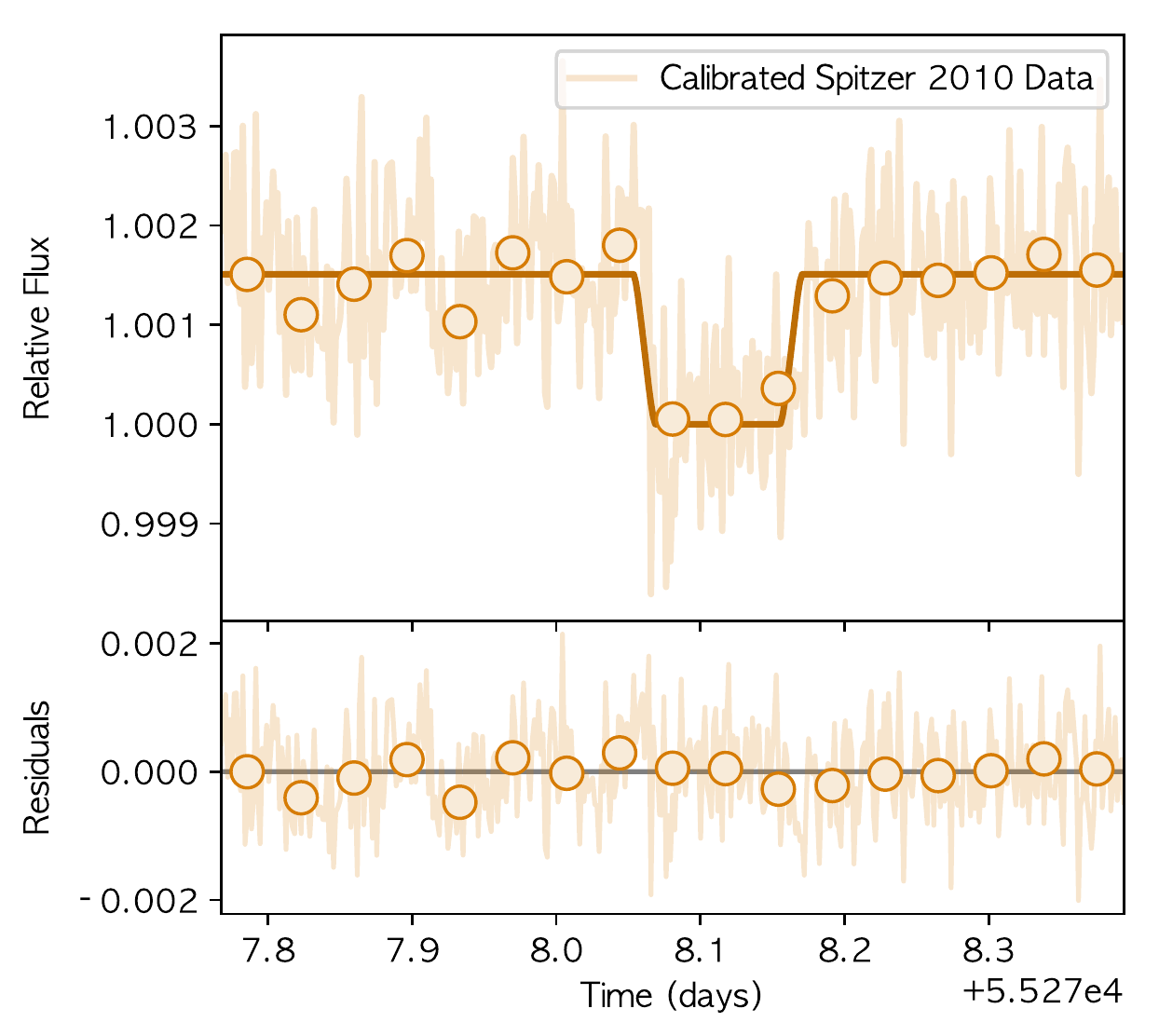}
\caption{SPCA analysis of the secondary eclipse obtained as part of the 2010 partial phase curve of XO-3b at 3.6 $\mu m$ with \emph{Spitzer}. The detrended data are connected with a pale orange line. The dark line represents the best-fit astrophysical model and the orange circles represent the binned calibrated photometry in 20 bins. The bottom panel shows the residuals of the SPCA fit. \label{fig:2010_eclipse}}
\end{figure}

\subsubsection{\SI{3.6}{\um} Secondary Eclipse Inconsistencies}

While the many repeat observations of the $4.5~\mu$m eclipse of XO-3b enable a precise and robust measurements \citep{2014ApJ...794..134W}, the planet has only been observed in 3.6~$\mu$m with \textit{Spitzer} two other times: one secondary eclipse in 2009 \citep[][PID 525]{2010ApJ...711..111M} and an unpublished partial phase curve obtained in 2010 (PI: P.\ Machalek, PID 60058). Our 1770$\pm$180 ppm eclipse depth is $5 \sigma$ discrepant with the 3.6 $\mu m$ eclipse depth of $1010 \pm 40$ ppm reported by \cite{2010ApJ...711..111M} taken during the cryogenic era \textit{Spitzer} data. Such significant discrepancies between cryogenic and warm \textit{Spitzer} eclipse depths are not unheard of \citep{hansen2014}, and we note that there is visible correlated noise in \cite{2010ApJ...711..111M}'s 3.6 $\mu$m residuals, hence their eclipse uncertainty is likely underestimated.

We investigate the 3.6 $\mu$m eclipse depths inconsistencies by analyzing the partial phase curve. Unfortunately, the 2010 observations are difficult to detrend because they only cover half an orbit and have large AOR breaks. Nonetheless, we were able to fit the secondary eclipse portion of the 2010 time series, shown in Figure \ref{fig:2010_eclipse}. We find an eclipse depth of $1520 \pm 130$ ppm, within $2 \sigma$ our fit to the 2013 full-orbit phase curve.

\cite{2010ApJ...711..111M}'s data were taken with a two-channel mode, i.e., two 2 s exposures at 3.6 $\mu$m for every 12 s exposure at 5.8 $\mu$m in order to avoid saturating in the shorter wavelength. As a result, the early eclipse has approximately 30\% the efficiency of the continuous observation mode used for the later phase curves. Hence, we elect to discard \cite{2010ApJ...711..111M}'s and the spoiled second eclipse in our full phase curve. When fitting the 2013 phase curve we experiment with using a Gaussian prior centered on the depth we obtained using the 2010 data. We find that the eclipse depth posteriors are consistent with or without the prior.


\subsubsection{Tentative \SI{3.6}{\um} EBM Fit}
While the average 4.5 $\mu$m \textit{Spitzer} temperature curve is consistent with the expected $T_{eq}$ of XO-3b, the 3.6 $\mu$m temperature curve is higher and does not intersect with the 4.5~$\mu$m temperature curve at any point in the orbit. In fact, the 3.6 $\mu$m brightness temperature curve is comparable to the planet's irradiation temperature, which means one of two scenarios: 1) a large excess flux or 2) leftover detector systematics. We favour the second scenario---detector systematics---but explore the implications of the model fit below for completeness.

The best-fit model gives an unexpectedly high $v_{\rm wind}$ of $12.1 ^{+3.0} _{-4.4}$  km/s, an order of magnitude faster than the equatorial wind speed obtained for the 4.5~$\mu$m fit. Such a large equatorial wind speeds are in general not physically possible. As highlighted in \cite{2018ApJ...853..133K} although wind speeds approaching or even slightly exceeding the speed of sound ($\sim$2 km/s) are possible in hot Jupiters, the development of shocks and shear instabilities in the atmosphere will naturally limit the maximal wind speeds at the atmospheric pressures being probed by our observations of XO-3b. The fit suggests a mixed layer down to $P_0 = 0.15 ^{+0.21} _{-0.11}$ bar, a short advective timescale of 0.2 hours and an average radiative timescale of $\sim 6$ hours. An internal energy source flux of $7.21 ^{+0.11} _{-0.17} \times 10^5$ W/m$^2$ is required to fit the 3.6 $\mu$m brightness temperature. This internal flux is \textbf{1.3 times} the average incident stellar flux. If this is interpreted as tidal heating, it would require a tidal quality factor of \textbf{$Q \sim 6 \times 10^3$}. Given the above issue with the 3.6~$\mu$m data, these results should be taken lightly.\\


\begin{figure}[!htpb]
\includegraphics[width=\linewidth]{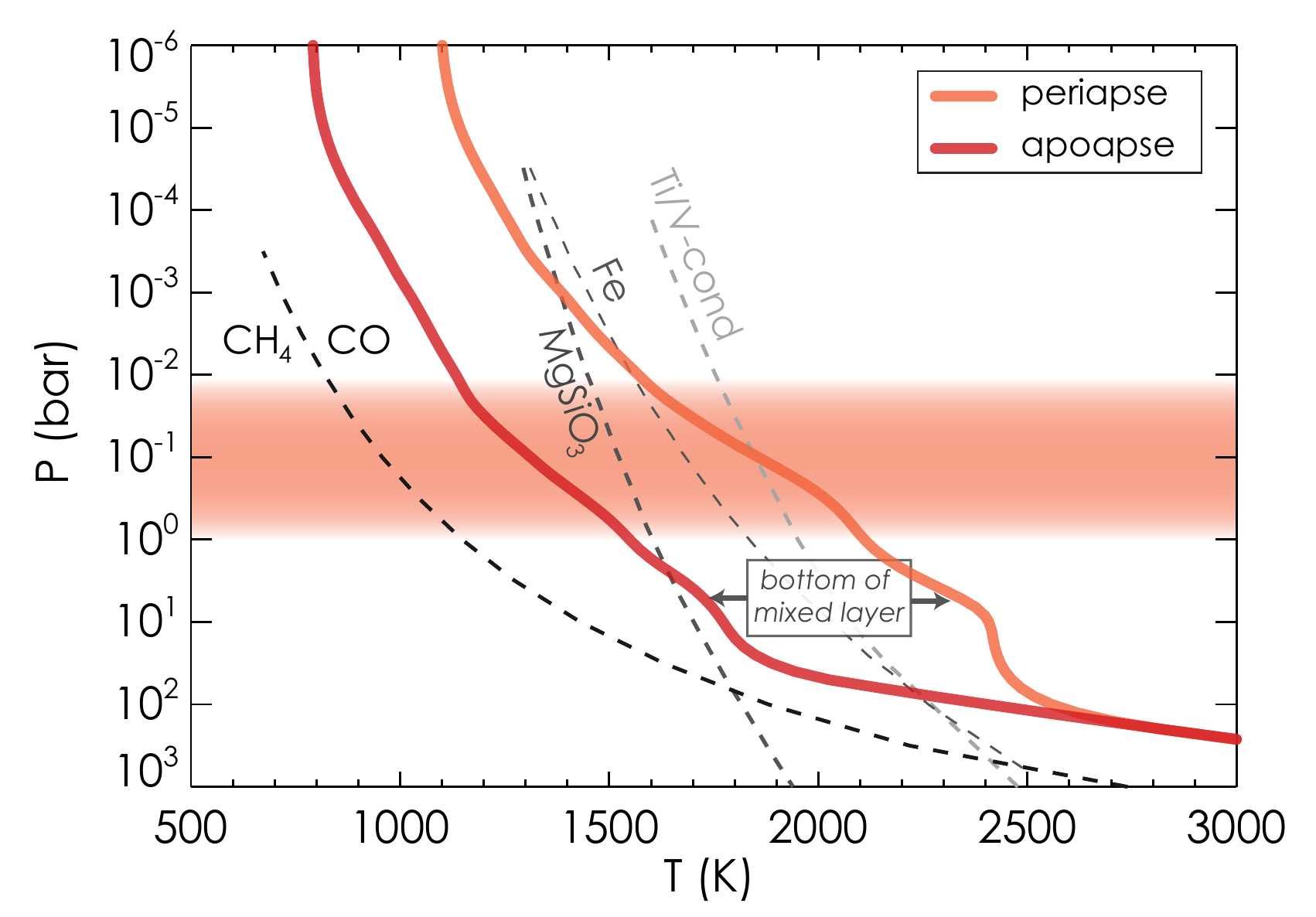}
\caption{1D atmospheric model at apoapse and periapse. The model predicts that incoming shortwave radiation from the star is deposited entirely at pressures $P<10$ bars above the isothermal region. An approximate infrared photosphere is indicated by the red swath from 0.01 to 1 bar. \label{fig:1D TP}}
\end{figure}

\begin{figure}[!htpb]
\includegraphics[width=\linewidth]{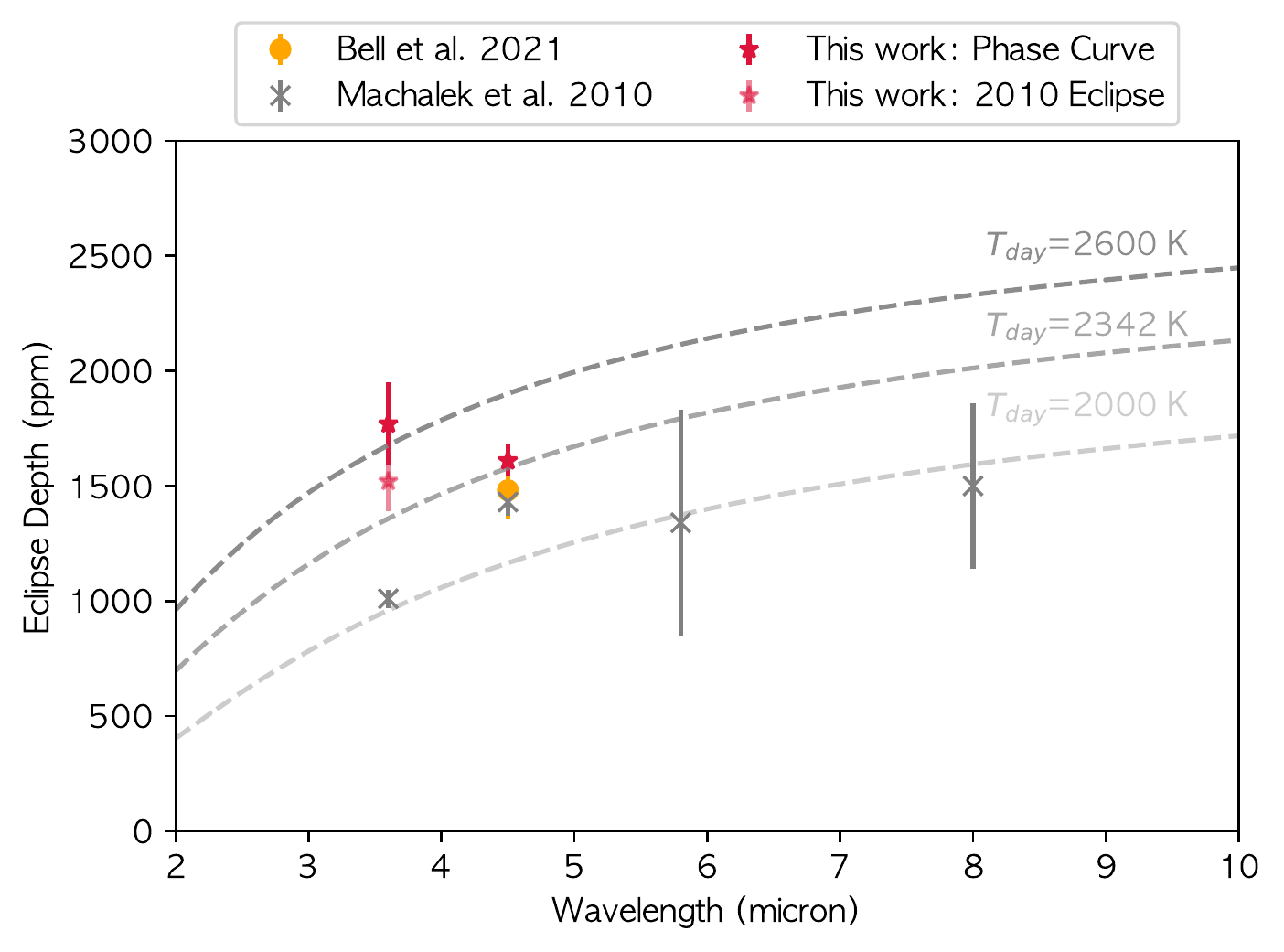}
\caption{Published \textit{Spitzer}/IRAC secondary eclipse depths are show along with our new eclipse depths. Gray dashed lines show notional eclipse spectra if the star and planet radiate as blackbodies with $T_{\rm eff, *} = 6885$ K and planetary dayside temperature of $T_{\rm day} = [2000, 2342, 2600]$ K. The deeper 3.6 $\mu$m eclipse depth in comparison to the 4.5 $\mu$m eclipse depth rules out the thermal inversion reported by \textcolor{blue}{Machalek et al. (2010)}: water vapour opacity dictates that the 3.6 $\mu$m photons originate from deeper in the atmosphere than those at 4.5 $\mu$m.  \label{fig:Emission Dayside}}
\end{figure}

\subsection{Theoretical Models of XO-3b's Atmosphere}

\subsubsection{Vertical Thermal Structure}

We present in Figure \ref{fig:1D TP} a planet-averaged 1D model at apoapse and periapse to look at the deposition of stellar energy as a function of pressure \citep{2002ApJ...568..335M, 2008ApJ...678.1419F}. The model uses the code, physics, and chemistry described in \citet{2008ApJ...678.1419F}. The specific entropy of the deep adiabat is relatively uncertain, here a value of the intrinsic flux (parameterized as $T_{\rm int}$) was set to 300 K a periapse, which sets the adiabat for the self-consistent model in radiative-convective equilibrium.  The temperatures in the deep part of the atmosphere are expected to be horizontally uniform.  Therefore, the $T_{\rm int}$ value for apoapse was iterated until the converged apoapse model fell upon the same deep adiabat, yielding $T_{\rm int}$=520 K. The planet's temperature should be homogenized at depth, but, $T_{\rm{int}}$ can't be thought of as constant in a simplified 1D modeling framework because the T-P profiles won't lie on the same adiabat due to limitations of a 1D model. We note that a more sophisticated solution with time-stepping atmospheric structure code has been developed to investigate the continuous atmospheric response to the variable incident flux that eccentric planets experience \citep{2021arXiv210508009M} that uses a different approach by fixing $T_{\rm int}$.

The 1D atmospheric model shown in Figure \ref{fig:1D TP} predicts that incoming shortwave light from the star has all been absorbed by $P<10$ bars. This is slightly deeper than the bottom of the mixed layer of $P_0=2.40 ^{+0.92} _{-0.16}$ bar inferred from our EBM fit to the $4.5~\mu$m data. We note that mixed layer here is a relevant model quantity in terms of explaining heat transport at observable levels, but it might not be accurate to extrapolate its meaning to the full depth of the circulation. In reality, because of the barotropic nature of the flow, winds will be strong throughout the observable atmosphere. A hotter interior would result in an adiabat at lower pressures and the isothermal region would take less room in pressure space. The infrared photosphere, on the other hand, is expected to be at the 0.01--1 bar altitude due to the atmosphere's greater IR opacity. In reality, for strong narrow absorption lines, photons could absorbed even down to $10^{-4}$ bars. We only expect winds to become significant at higher altitude than the deposition layer, where horizontal temperature gradients are greater; we therefore expect the mixed layer to lie above the shortwave deposition depth. 

By comparing the simulated temperature-pressure profiles with condensation and molecular transition curves in Figure \ref{fig:1D TP}, we expect that CO is the main carbon carrier on the planet throughout its orbit. Clouds of TiO/VO, FeSiO$_{\rm 3}$ and MgSiO$_{\rm 3}$ might be expected to form at or above the IR photosphere  when the planet is near periapse, but would be too deep to affect the emergent spectrum at apoapse.  Hence the emergent spectrum of the planet near periapse---possibly including the secondary eclipse---could be affected by the presence of clouds above the notional clear-sky photosphere. There is no evidence of such clouds in the planet's eclipse spectrum.
The \textit{Spitzer} eclipse depth measurements are shown in Figure \ref{fig:Emission Dayside}: a deeper 3.6 $\mu$m eclipse depth disfavors the dayside thermal inversion reported by \cite{2010ApJ...711..111M}.
Unfortunately, our analysis is inconclusive as to temperature inversions for the rest of the orbit due to systematics spoiling our 3.6 $\mu$m phase curve. Further investigations with JWST at different orbital phases could provide a better understanding of XO-3b's atmospheric thermal structure and its response to the changing incident stellar flux.

\begin{figure*}[!hptb]
\includegraphics[width=.495\linewidth]{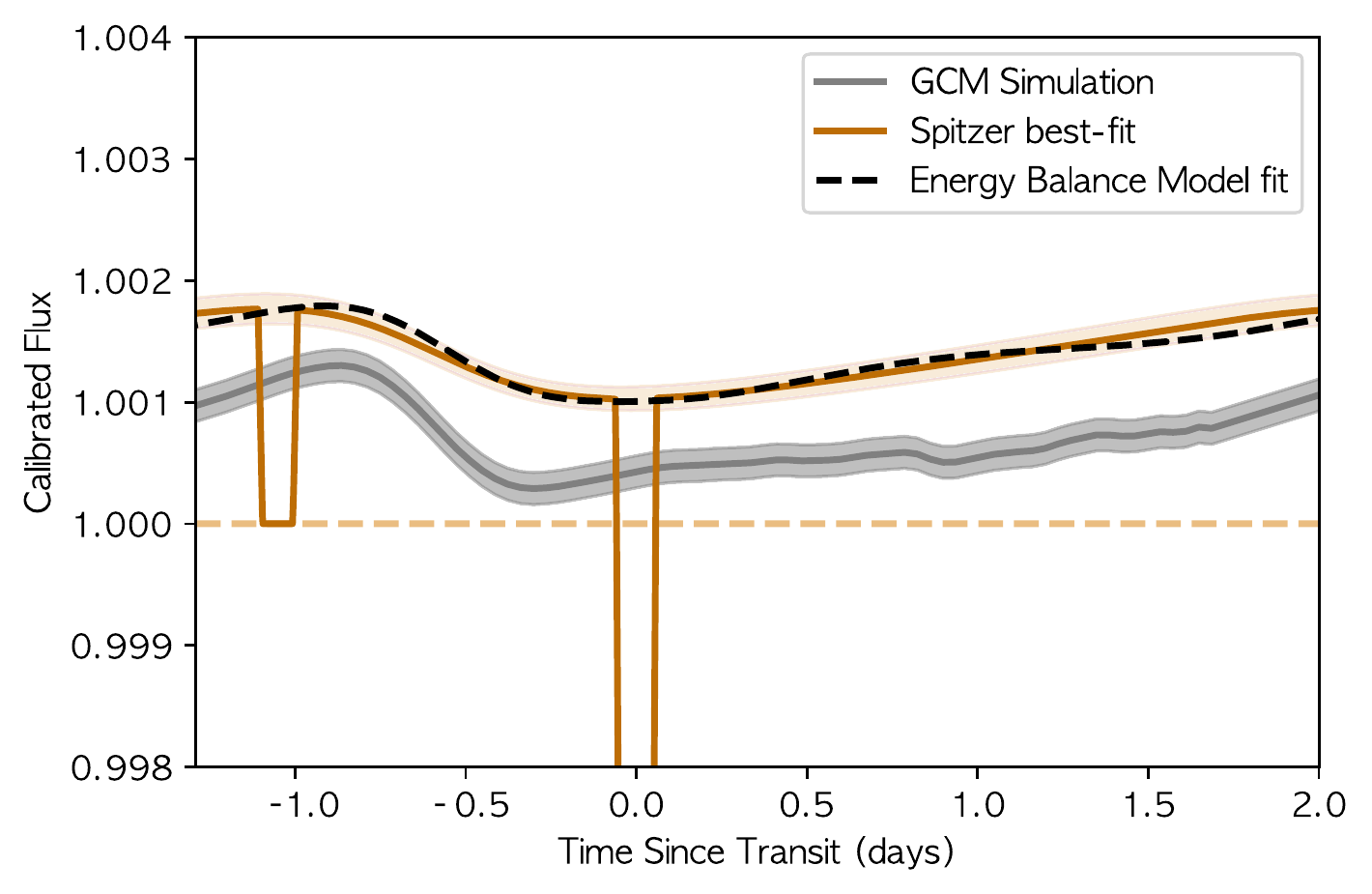}\includegraphics[width=.495\linewidth]{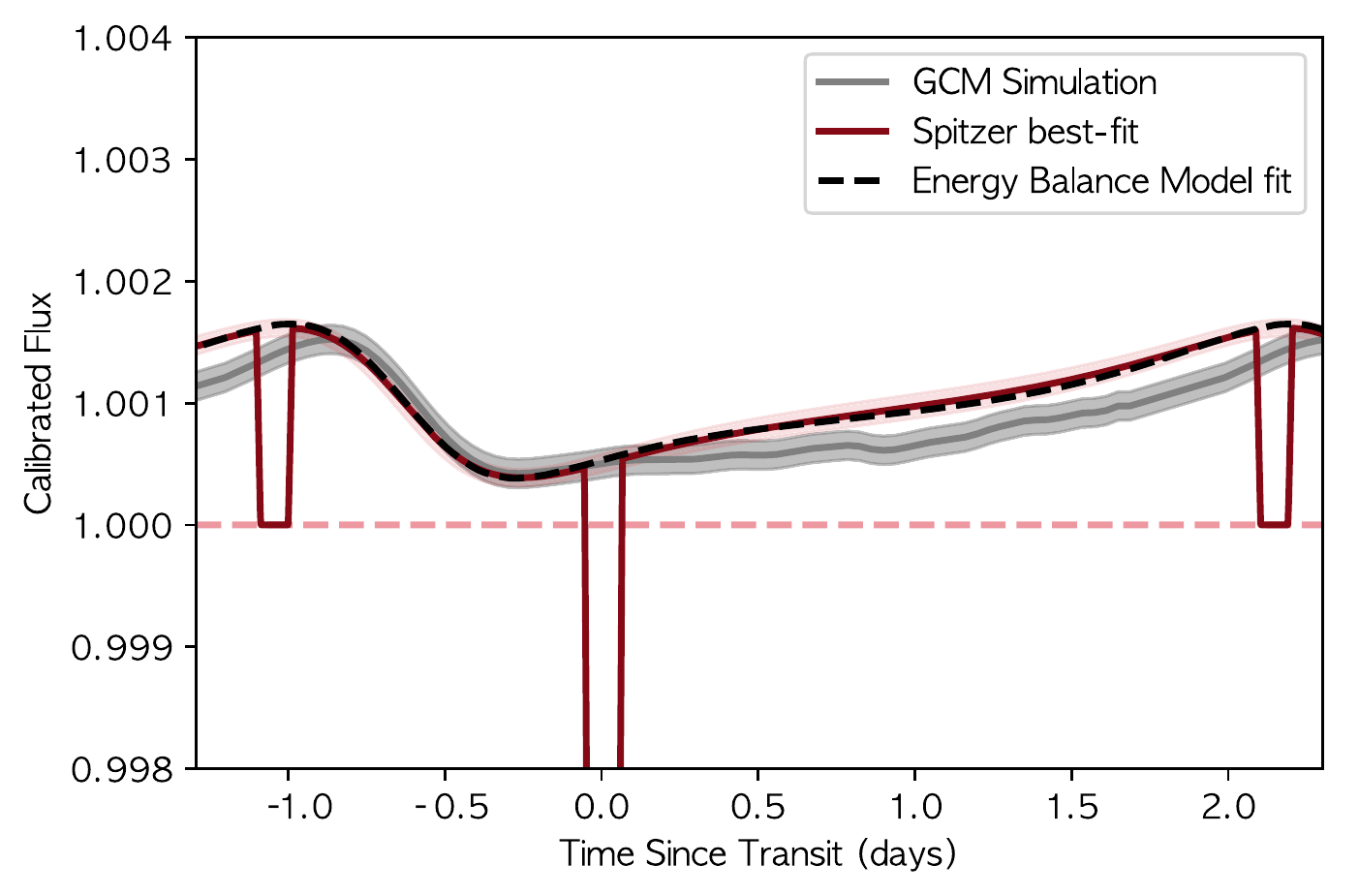}
\caption{{\bf General} circulation model (GCM) thermal phase curve predictions for the 3.6$~\mu$m (\textit{left}) and 4.5$~\mu$m (\textit{right}) Spitzer band passes. The coloured swath represent the 1$\sigma$ uncertainties of the best-fit \textit{Spitzer} phase curves. The grey swath represents the magnitude of the orbit-to-orbit variability seen in the GCM. The large discrepancy between the 3.6$~\mu$m \textit{Spitzer} phase curve and GCM predictions is likely due to issues with the observations. At 4.5$~\mu$m, our model correctly predicts the cooling rate of the planet between eclipse and transit, but underestimates its heating rate between transit and eclipse.The dashed black lines represents the EBM fit to each Spitzer phase curve. \label{fig:GCM-pc}}
\end{figure*}

\begin{figure*}[!hptb]
\includegraphics[width=.495\linewidth]{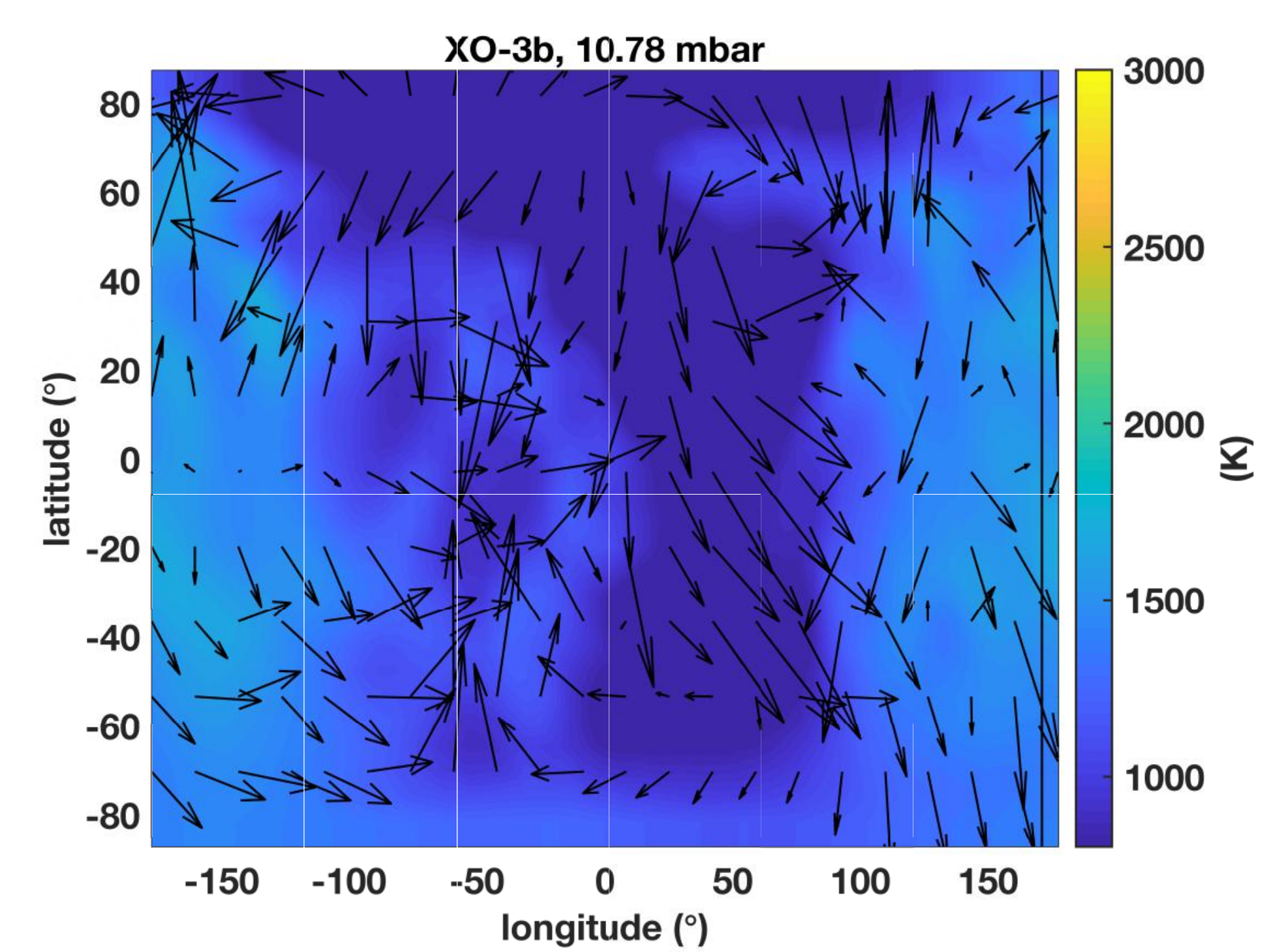}\includegraphics[width=.495\linewidth]{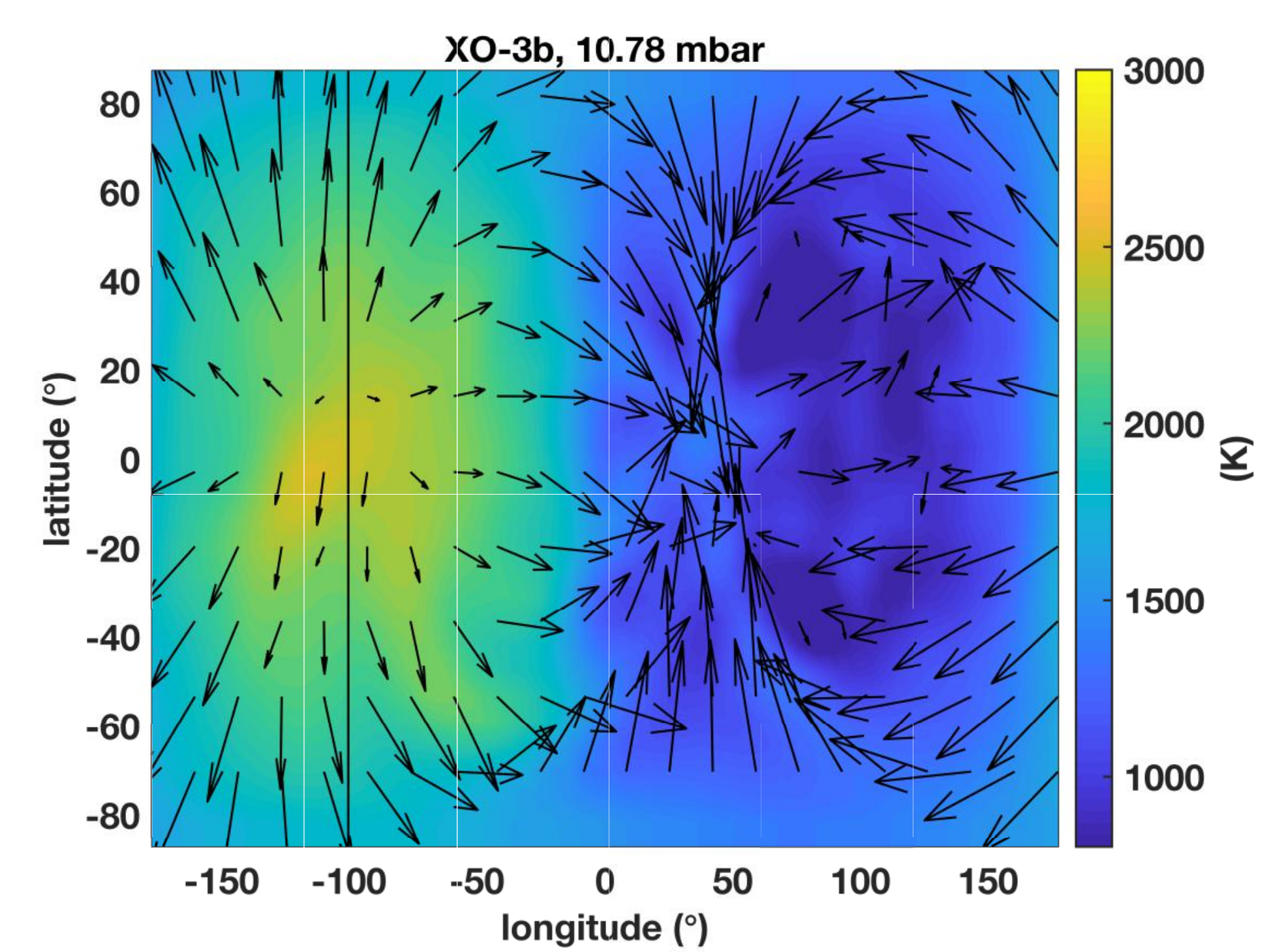}
\includegraphics[width=.495\linewidth]{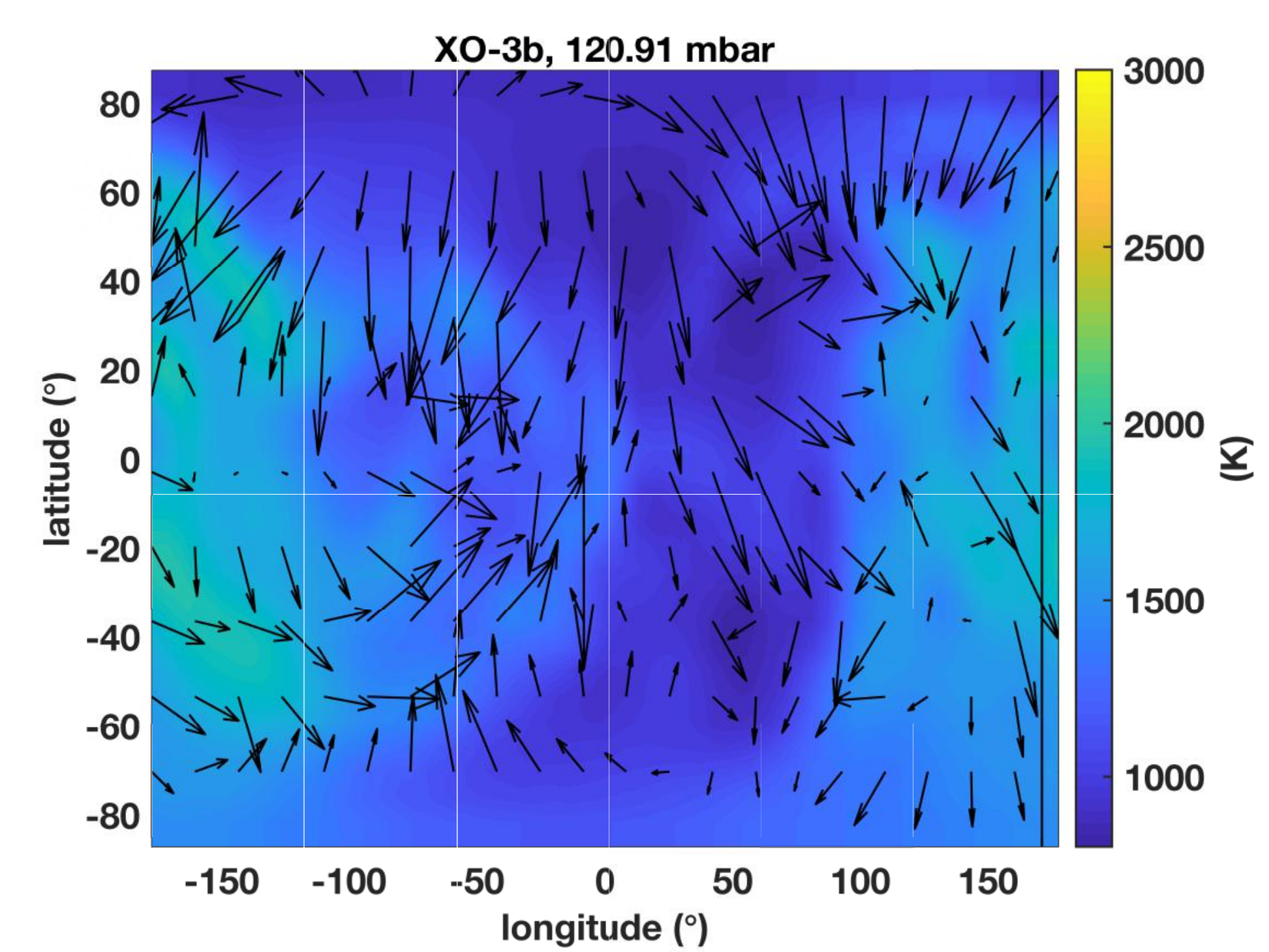}\includegraphics[width=.495\linewidth]{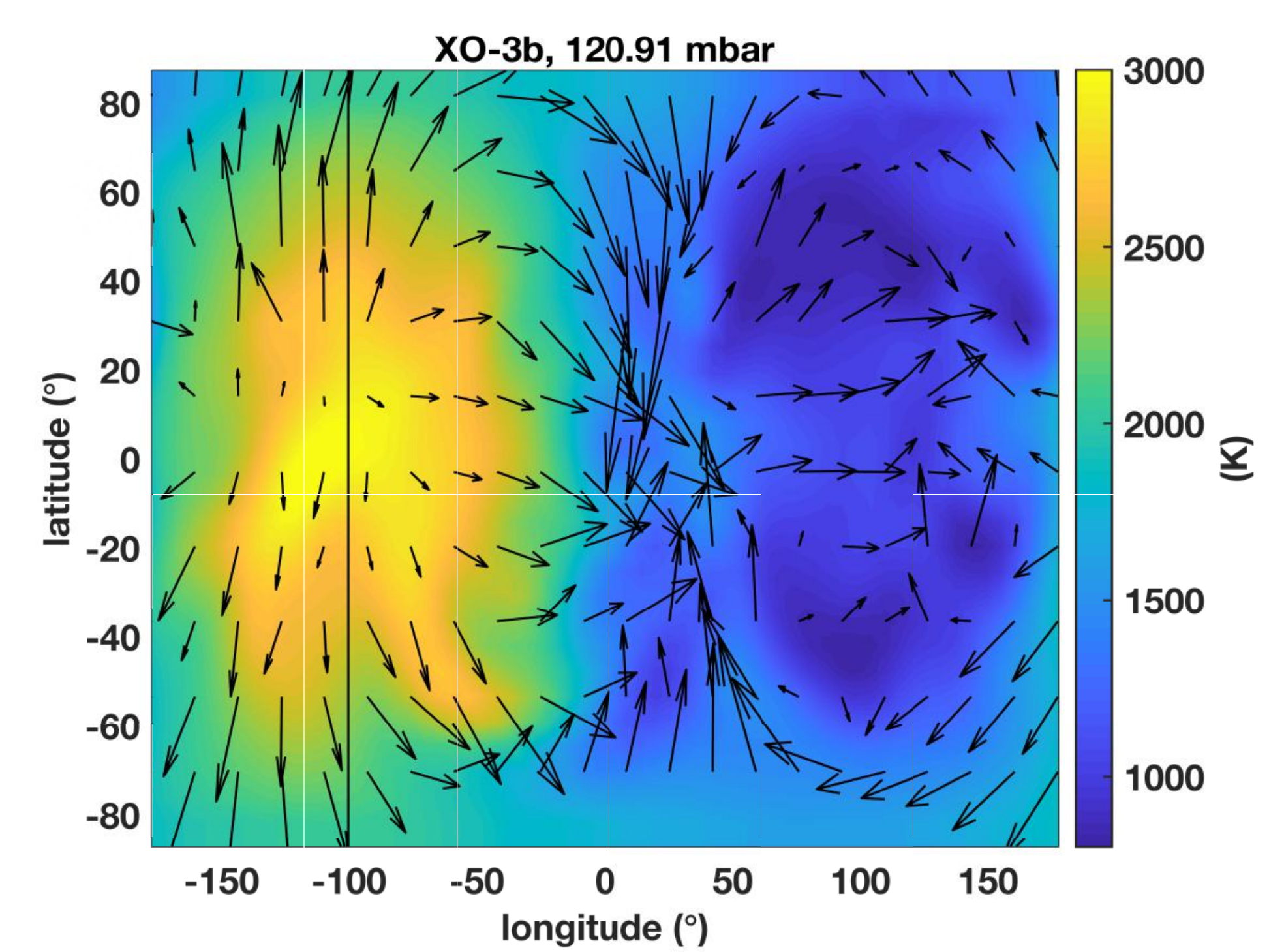}
\caption{Temperature (colorscale) and winds (arrows) at the $\approx$ 10 mbar and $\approx$ 120 mbar level of our 1$\times$ solar model at snapshots corresponding to apoapse (left) and periapse (right). The substellar longitude is indicated by the solid vertical line. At periapse, the model exhibits a large temperature contrast between the dayside and nightside of the planet. At apoapse, the temperature gradient is attenuate and the predominant wind zonal and meridional flows are suppressed.} \label{fig:GCM_v2}
\end{figure*}

\subsubsection{3D General Circulation Model}

We present three-dimensional atmospheric circulation models of XO-3b using the SPARC/MITgcm \citep{Showman2009}. The SPARC/MITgcm couples the MITgcm. a three-dimensional (3D) general circulation model (GCM) \citep[GCM;][]{Adcroft2004} with a two-stream adaptation of a multi-stream radiative transfer code \cite{Marley1999}. The MITgcm solves the primitive equations using the finite-volume method over a cubed sphere grid. The radiative transfer code solves the two-stream radiative transfer equations, and employs the correlated-$k$ method to solve for upward/downward fluxes and heating/cooling rates through the atmosphere \cite[e.g.,][]{Goody1989,Marley1999}. The correlated-$k$ method retains most of the accuracy of full line-by-line calculations, while drastically increasing computational efficiency. The SPARC/MITgcm has been applied to a range of exoplanets and brown dwarfs \cite[e.g.,][]{Showman2009,Kataria2014,Lewis2014,Parmentier2018,Steinrueck2019}.

In this model we adopt a horizontal resolution of 32$\times$64 in latitude and longitude, and a vertical resolution of 40 layers evenly spaced in log pressure from 200 bars at the bottom boundary to 200 microbars at the top. Given that XO-3b is on an eccentric orbit, we assume the planet is ``pseudo-synchronously" rotating, i.e., that the planet’s tidal interactions with the star force a single side of the planet to approximately face the star every periapse passage. We estimate the planet’s rotation rate following the \cite{1981A&A....99..126H} formulation for binary stars, $T_{\rm rot}=1.852\times10^5$ seconds (or approximately 2 days). We assume a Solar atmospheric composition without TiO/VO (whose opacities are used to produce a thermal inversion). Given the high gravity and eccentricity of the planet, the GCM was run for 63 Earth days. Despite this short run time, this amounts to approximately 21 orbits of XO-3b, sufficient time for the model to have converged. 

Unlike hot Jupiters on circular orbits, eccentric hot Jupiters allow us to investigate the atmospheric response to time-varying incident flux. We use our 3D simulations to inspect wind patterns and temperature gradients at apoapse and periapse at pressures of 10 and 100 mbar (Figure \ref{fig:GCM_v2}). At periapse, the GCM predicts a large temperature contrast between the dayside and nightside of the planet with large zonal (east-west) and meridional (north-south) flows from the substellar point to the limbs and antistellar point. Conversely, at apoapse, the atmosphere is comparatively quiescent, with low temperature contrasts from dayside to nightside, and unorganized flow. This behavior is in broad agreement with previous GCMs of highly eccentric exoplanets, including HD~80606b \citep{Lewis2017} and HAT-P-2b \citep{Lewis2014}. 

Comparing GCM predictions to our \textit{Spitzer} phase curves in Figure \ref{fig:GCM-pc}, we find that the planetary 3.6 $\mu$m flux is greatly underestimated throughout the orbit. As noted in section 5.3, this discrepancy is likely due to issues with the 3.6 $\mu$m observations. However, it is surprising that the 3.6 $\mu$m eclipse depth is also underestimated. Assuming the absence of the formation of a strong thermal inversion in the dayside atmosphere of XO-3b, the relative flux from the planet at 3.6 $\mu$m vs.\ 4.5 $\mu$m is a strong function of the pressures and hence atmospheric temperatures being probed by each channel. As the GCM assumes instantaneous equilibrium chemistry in the atmosphere it cannot capture possible disequilibrium processes that may affect abundances of key species such as $\mathrm{CH_4}$ and CO that are strong absorbers in the 3.6 and 4.5 $\mu$m \textit{Spitzer} bandpasses respectively. \cite{Visscher2012} highlights that for eccentric hot Jupiters like XO-3b orbit induced thermal quenching can produce a significant reduction in the abundance of $\mathrm{CH_4}$ in the planet’s atmosphere throughout its orbit. Such a scenario would naturally allow the 3.6 $\mu$m channel to probe deeper into XO-3b’s atmosphere resulting in a deeper than expected secondary eclipse depth in that channel. 

The numerical model is able to correctly predict the amplitude of the 4.5 $\mu$m phase curve, although the peak and trough occurs slightly later than the \textit{Spitzer's} data. The model seems to adequately predict the cooling timescale of the planet at 4.5 $\mu$m, but underestimates the heating timescale where the discrepancy is more apparent after transit. Although predictions from the GCM match well with the flux measured from XO-b from eclipse through periastron and into transit, especially at 4.5 $\mu$m, the GCM predicts a significantly shallower increase in the planetary flux between the transit and eclipse events. As highlighted in other studies of eccentric hot Jupiters such as HAT-P-2b \citep[e.g,][]{Lewis2013,Lewis2014} and HD80606b \citep[e.g.,][]{deWit2016}, the assumption of a ``pseudo-synchronous'' rotation rate for hot Jupiters on eccentric orbits can result in inconsistencies between model predictions and observations. Near periastron passage, the thermal structure of the planet is dominated by the intense transient heating that results in the theoretical flux from the planet to be fairly insensitive to the assumed rotation rate. However away from periaston passage the assumed rotation rate plays a stronger role in shaping the phase dependent flux from the planet (see discussion in \citealt{Lewis2014} in the context of HAT-P-2b and \citealt{deWit2016} in the context of HD80606b). A rotation rate that is slower than the assumed pseudo-synchronous rotation rate would result in the cooler hemisphere of XO-3b being projected toward an earth observer for more the time between the transit and secondary eclipse event that would mimic a slower than expected heating rate for the planet.

The energy balance model allows $P_0$, $A_b$, $E_{int}$ and $V_{\rm wind}$ to be free-parameters, constant across the planet and throughout the orbit. In contrast, these quantities are spatially and temporally variable in the GCM and the local pressure levels of absorption and re-emission, Bond Albedo and winds are computed self-consistently assuming equilibrium chemistry. Additionally, the EBM is compared with each \textit{Spitzer} phase curve separately while the simulated GCM phase curves are derived from the same simulation. The internal heat, $E_{int}$, is a free parameter that is explored in the EBM but not in the GCM. Given the atmospheric temperature expected for XO-3b and assumption of chemical equilibrium, the 3.6 micron photosphere will generally be located at deeper pressures (~400 mbar) compared to the 4.5 $\mu$m photosphere (~100 mbar). Therefore, increasing internal heat in the GCM could serve to increase the temperature at depth and provide a better prediction to the 4.5 $\mu$m phase curve. The discrepancy between the 3.6 $\mu$m phase curve and the GCM prediction could also be due instrumental issues with the 3.6 $\mu$m channel. However, the GCM also under-predicts the robust 3.6 $\mu$m eclipse depth and instrumental effects are unlikely to be the cause for this difference.

\subsection{Possible Inflated Radius of XO-3b}

\begin{figure}[!htpb]
\includegraphics[width=\linewidth]{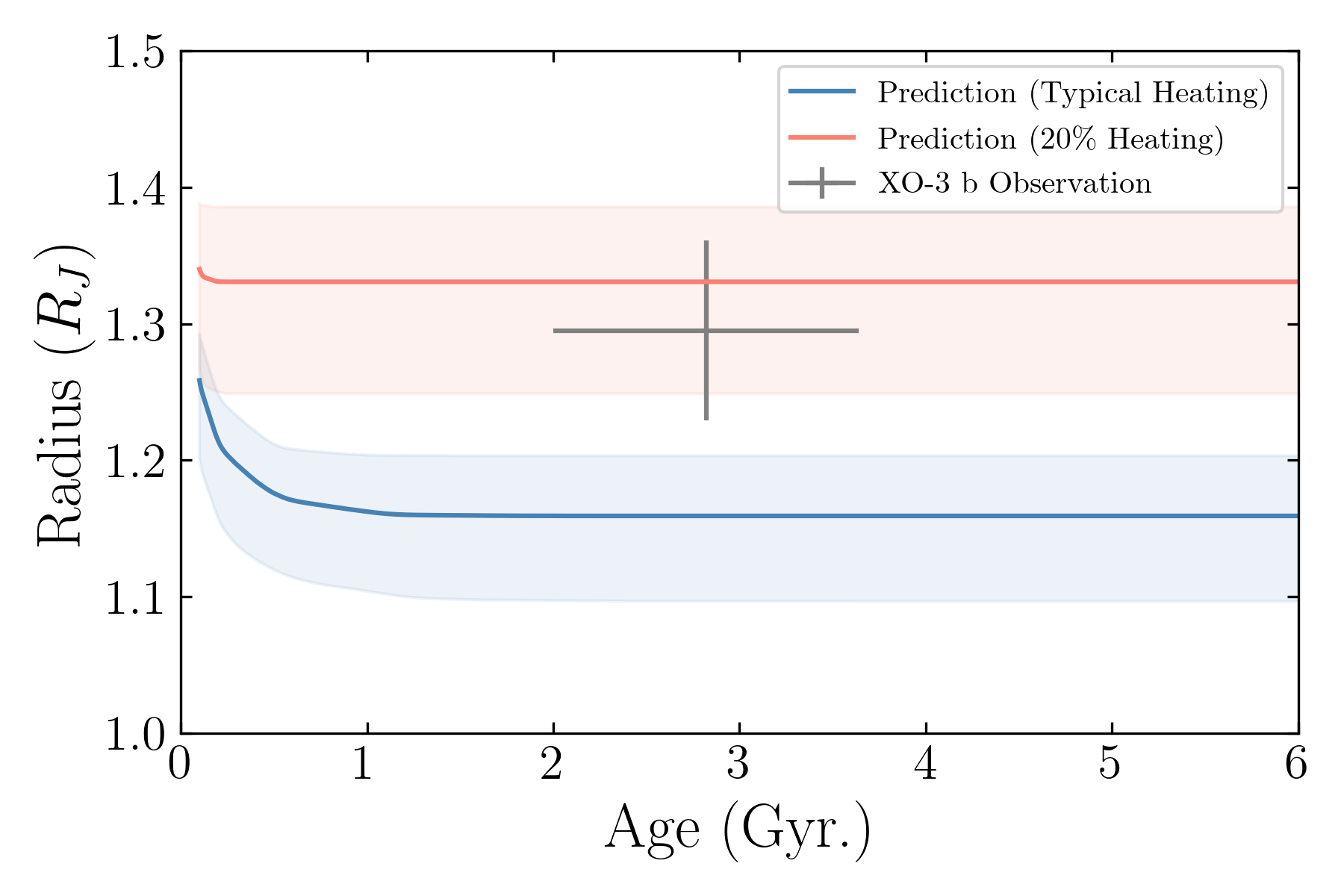}
\caption{Theoretical planet radius vs age for XO-3b with typical heating (blue) and with additional heating (red) compared to XO-3b's observed radius. The shaded swatchs represent model uncertainties, dominated by the compositional uncertainties. \label{fig:Evolution Model}}
\end{figure}

Given the significant difference between the radius and mass of XO-3 reported by \cite{2017AJ....153..136S} and previously reported parameters, we re-determined these parameters by updating the stellar $T_{\rm eff} = 6759 \pm 79$~K from the latest PASTEL spectroscopic catalog \citep{2020yCat....102029S} and the parallax to the {\it Gaia\/} EDR3 value \citep{2021A&A...649A...1G}, then applied all of the other empirical parameters and calculations described in \citet{2017AJ....153..136S} with a correction described in \cite{2018ApJ...862...61S}. We find a stellar mass and radius of $M_* = 1.21 \pm 0.15$ $M_{\rm{\odot}}$ and $R_* = 1.407 \pm 0.038 R_{\rm{\odot}}$ and a planetary mass and radius of $M_p = 11.79 \pm 0.98$ $M_{\rm J}$ and $R_p = 1.295 \pm 0.066 R_{\rm J}$.

To contextualize these new constraints, we compare them with planetary interior structure models based on \citet{2018AJ....155..214T}.  These are 1-D evolution models that solve the equations of hydrostatic equilibrium, mass conservation, and the relevant equations of state.  The most important free parameters are the bulk metallicity $Z_p$ and the anomalous heating, which we parametrize as a fraction of the incident flux.  Using heating values fitted from the observed hot Jupiter population \citep{2018AJ....155..214T} and a distribution of bulk metallicities inferred from the warm Jupiter population \citep{Thorngren2016}, we predict a range of expected radii for the observed mass and flux.

Figure \ref{fig:Evolution Model} shows this radius plotted against age, and suggests that the observed radius of XO-3b is about 2$\sigma$ larger than expected.  Interestingly the expected radius of XO-3b is consistent with the observed radius of the planet if we use a high 20\% insolation interior heating.  Such \textbf{a} relationship between the internal heating and planetary radius has been previously proposed for KELT-1b, a 27-$M_{\rm J}$ brown dwarf companion \citep{2021A&A...648A..71V}. Given the observational uncertainties, this means that we are either  measuring the radius at 2 sigma above the true value or the planet is inflated beyond the level expected for its time-averaged flux \citep{2019ApJ...884L...6T}.

If we interpret the radius of the planet as being linked to internal heating, then there are many possible candidate sources of heat. First, the planet is expected to experience tidal heating and it is possible we are catching the planet in a few Myr window during which it is rapidly circularizing \citep[][]{2002ApJ...573..829M, 2009ApJ...700.1921I, 2019ApJ...886...72M}.  Since the observed mass of the planet is less than 1$\sigma$ below the deuterium limit, it is possible that the true mass of the planet is slightly above the deuterium limit and it could take gigayears to finish burning \citep{2011ApJ...727...57S, 2020A&A...637A..38P}.  Even if somewhat below this limit, it is likely that at least some of the deuterium in the planet has or will be burned \citep{Spiegel2011, Bodenheimer2013}; the main issue therefore is whether the heating is sufficient to explain the radius at this age, but that is a difficult question outside the scope of this work.  Another possibility is that the planet could be experiencing a Cassini State 2 with high obliquity that would increase the tidal dissipation of the planet \citep{2007ApJ...665..754F, 2019AJ....158..108A}. Thermal tides, advection of potential temperatures, and Ohmic dissipation are proposed mechanisms for the general radius inflation problem that may also play an important role on XO-3b \citep[e.g.][]{2013arXiv1304.4121S, 2017ApJ...841...30T, 2018AJ....155..214T}. Ultimately, unusually hot interior would likely be explained by a combination of the usual hot Jupiter inflation effect operating in conjunction with one or more of these other, less common heat sources.

\section{Summary and Conclusion} \label{sec:conclu}

We presented the analysis of new \textit{Spitzer}/IRAC observations of the curious XO-3 system harbouring a massive $M_p = 11.79 \pm 0.98$ $M_{\rm{jup}}$ inflated hot Jupiter with $R_p = 1.295 \pm 0.066 R_{\rm{jup}}$ on a 3.2 day orbit with an orbital eccentricity of $e=0.2769$. The full-orbit 3.6 $\mu$m and 4.5 $\mu$m \textit{Spitzer}/IRAC phase curves of XO-3b yield a secondary eclipse depths of 1770 $\pm$ 180 ppm and 1610 $\pm$ 70 ppm at 3.6 $\mu$m and 4.5 $\mu$m, respectively. From the secondary eclipse portion of the 3.6 $\mu$m partial phase curve of XO-3b obtained in 2010 (PI: P. Machalek, PID 60058), we retrieve an eclipse depth of 1520 $\pm$ 130 ppm which agrees with our more recent phase curve observations, but is 5$\sigma$ discrepant with \cite{2010ApJ...711..111M}'s results. Our observations therefore suggest no evidence for the thermal inversion on the dayside at secondary eclipse proposed by \cite{2010ApJ...711..111M}. The discrepancy is likely due to the less efficient observing mode used by \cite{2010ApJ...711..111M} and resulting systematics.

Unfortunately, detector systematics are difficult to decorrelate from our 3.6 $\mu$m phase curve. Nonetheless, we compare our reliable 4.5 $\mu$m phase curve observations to multiple atmospheric models to constraint the radiative and advective properties of XO-3b. We use an energy balance model, assuming the \cite{1981A&A....99..126H} prescription for pseudosynchronous rotation rate,
to fit the more reliable 4.5 $\mu$m observations and find a Bond albedo of $A_b=0.106 ^{+0.008}_{-0.106}$ best-fits our data. We also estimates an average equatorial wind speed $v_{\rm wind}$ of $3.13 ^{+0.26} _{-0.83}$ km/s, in agreement with the $\sim$~2.5 km/s equatorial wind speeds predicted near periastron by a general circulation model. Our energy balance model fit suggest that the mixed layer of the atmosphere on a planet-averaged extends down to $P_0 = 2.40 ^{+0.92} _{-0.16}$ bars which is consistent with our 1D radiative transfer model that predicts shortwave light absorbed at deeper pressures. We also compare our phase curves with predictions from a GCM and find good agreement at 4.5 $\mu$m and large discrepancies at 3.6 $\mu$m. While the disagreement could be due to detector systematics spoiling the 3.6 $\mu$m phase curve, it's unlikely to be culprit for the difference in \textit{Spitzer}-measured and GCM-predicted 3.6 $\mu$m eclipse depths. Planetary evolution models suggest that XO-3b is unusually large for its mass. Interestingly, additional heating equivalent to 20\% insolation could explain its observed radius. If our results are interpreted as internal heat, the cryptic source of heating could be deuterium burning or tidal dissipation due to the orbital eccentricity or the high planetary obliquity.

Better characterization of stellar properties resulting in stringent constraints on the planet's mass would allow us to determine if the radius of XO-3b is really unusual. Further investigations with the James Webb Space Telescope would enable a search for clouds and could better constrain the presence of a temperature inversion at orbital phases other than at superior conjunction. Phase curve observations at other wavelengths can also better constraint the planetary flux and hence cryptic heating. Along with, HD 80606b, a giant planet with an orbital eccentricity of 0.93, gas giants with moderate orbital eccentricity, such as XO-3b and HAT-P-2b, offer a unique opportunity to characterize the gas giants at different stages of planet migration and help constrain planetary evolution theories.



\acknowledgments

L.D. acknowledges support in part through the Technologies for Exo-Planetary Science (TEPS) PhD Fellowship, and the Natural Sciences and Engineering Research Council of Canada (NSERC)'s Postgraduate Scholarships-Doctoral Fellowship. This work is based on archival data obtained with the Spitzer Space Telescope, which is operated by the Jet Propulsion Laboratory, California Institute of Technology under a contract with NASA. T.J.B.~acknowledges support from the McGill Space Institute Graduate Fellowship, the Natural Sciences and Engineering Research Council of Canada's Postgraduate Scholarships-Doctoral Fellowship, and from the Fonds de recherche du Qu\'ebec -- Nature et technologies through the Centre de recherche en astrophysique du Qu\'ebec.

\facilities{Spitzer Space Telescope (IRAC)}
\software{\texttt{astropy} \citep{2013A&A...558A..33A}, \texttt{emcee} \citep{2013PASP..125..306F}, \texttt{batman} \citep{2015PASP..127.1161K}, \texttt{SPCA} \citep{2018NatAs...2..220D, 2021MNRAS.504.3316B}, \texttt{Bell\_EBM} \citep{2018ApJ...857L..20B}}

\newpage
\pagebreak
\appendix

\section{Tidal Heating}\label{sec:tides}

Time-dependent tidal distortion of a body leads to internal heating \citep{1978Icar...36..245P, 1979Sci...203..892P, 2004AJ....128..484W}. Hence, eccentricity and obliquity tides have been proposed as the missing energy source to explain the anomalously large radius of some hot Jupiters. We estimate the rate of energy dissipation using the formalism described in \cite{2007A&A...462L...5L} which takes into account the effect of synchronous rotation:
\begin{equation}
    \frac{dE_{\rm{tides}}}{dt} = 2K \left[ N_a(e) - \frac{N^2(e)}{\Omega (e)} \frac{2x^2}{1+x^2}\right]
\end{equation}

\noindent where $\Omega(e)$, $N(e)$, and $N_a(e)$ are functions of the orbital eccentricity and are defined as:
\begin{equation}
 \Omega(e)=\frac{1+3 e^{2}+\frac{3}{8} e^{4}}{\left(1-e^{2}\right)^{9 / 2}},
\end{equation}
\begin{equation}
  N(e)=\frac{1+\frac{15}{2} e^{2}+\frac{45}{8} e^{4}+\frac{5}{16} e^{6}}{\left(1-e^{2}\right)^{6}},
\end{equation}
\begin{equation}
    N_{a}(e)=\frac{\left(1+ \frac{31}{2} e^{2} +\frac{255}{8} e^{4} +\frac{185}{16} e^{6} +\frac{25}{64} e^{8} \right)}{(1-e^2)^{15/2}}.
\end{equation}

\noindent and where $x = \cos \epsilon$, $\epsilon$ is the planet's obliquity (the angle between the equatorial and orbital planes) and
\begin{equation}
    K = \frac{3}{2} \frac{k_2}{Q_n} \left( \frac{GM_p^2}{R_p}\right) \left(\frac{M_*}{M_p}\right)^2 \left( \frac{R_p}{a} \right)^6n
\end{equation}

\noindent where $k_2$ is the planet's potential Love number of degree 2, $Q_n$ is the planet's annual tidal quality factor, $G$ is the gravitational constant, and $n$ is the planet's mean motion which is approximately $\sqrt{GM_*/a^3}$. Assuming that XO-3b has a zero obliquity and Jupiter's tidal Love number $k_2=0.565\pm0.006$ \citep{2020GeoRL..4786572D}, we find that a tidal quality factor of $Q \sim 6 \times 10^3$ is required to explain the excess flux of $1.3 \times$ insolation inferred for the 3.6~$\mu$m phase curve. 

\bibliography{xo3b.bib}{}
\bibliographystyle{aasjournal}

\end{document}